\documentclass[twoside,showpacs,superscriptaddress,groupedaddress,singlecolumn]{revtex4}
\usepackage{hyperref}
\usepackage{color}
\usepackage{graphicx}
\usepackage{subfig}
\usepackage{epstopdf}
\usepackage{soul}
\usepackage{float}

\makeatletter
\@ifundefined{textcolor}{}
{
 \definecolor{BLACK}{gray}{0}
 \definecolor{WHITE}{gray}{1}
 \definecolor{RED}{rgb}{1,0,0}
 \definecolor{GREEN}{rgb}{0,1,0}
 \definecolor{BLUE}{rgb}{0,0,1}
 \definecolor{CYAN}{cmyk}{1,0,0,0}
 \definecolor{amber(sae/ece)}{rgb}{1.0, 0.49, 0.0}
 \definecolor{YELLOW}{cmyk}{0,0,1,0}
}

\def\be{\begin{equation}}
\def\ee{\end{equation}}
\def\ba{\begin{eqnarray}}
\def\ea{\end{eqnarray}}

\newcommand{\spp}{{s_{++}}}
\newcommand{\spm}{{s_{\pm}}}

\makeatother

\begin{document}

\title{Quasiparticle interference in multiband superconductors with strong coupling.}

\author{A. Dutt}
\affiliation{Faculty of Science and Technology and MESA+ Institute of Nanotechnology,
University of Twente, 7500 AE Enschede, The Netherlands}

\author{A.A.~Golubov}
\affiliation{Faculty of Science and Technology and MESA+ Institute of Nanotechnology,
University of Twente, 7500 AE Enschede, The Netherlands}
\affiliation{Moscow Institute of Physics and Technology, Dolgoprudnyi, Russia}
\author{O.V.\ Dolgov}

\affiliation{Max-Planck-Institut FKF, D-70569 Stuttgart, Germany}
\affiliation{P.N. Lebedev Physical Institute,Russian Academy of Science, Moscow,Russia}
\author{D.V.\ Efremov}

\affiliation{Leibniz-Institut f\"{u}r Festk\"{o}rper- und Werkstoffforschung Dresden,
Dresden, Germany}

\begin{abstract}
We develop a theory of the quasiparticle interference (QPI) in multiband
superconductors based on strong-coupling Eliashberg approach within the Born approximation. In the framework of this theory, we study  dependencies of the QPI response function
in the multiband superconductors with nodeless s-wave superconductive order  parameter. We pay a special attention to the difference of the quasiparticle scattering between the bands having the same and opposite signs of the order parameter. We show that, at the momentum values close to the momentum transfer between two bands, the energy dependence of the quasiparticle interference response function has three singularities. Two of these correspond
to the values of the gap functions and the third one depends on
both the gaps and the transfer momentum. We argue that only the singularity near the smallest band gap may be used as an universal tool to distinguish between $s_{++}$ and $s_\pm$ order parameters. 
The robustness of the sign of the response
function peak near the smaller gap value, irrespective of the change in
parameters, in both the symmetry cases is a promising feature that can be
harnessed experimentally.
\end{abstract}

\maketitle

\section{ Introduction}

In recent decades, a number of new materials such as cuprates, magnesium
diboride, chalcogenides and iron pnictides with a high critical temperature have been found. {\cite%
{Hosono2008,Mazin02,Stewart11,Paglione10,Johnston10,Hosono15}}
This generated numerous proposals for the mechanisms of superconductivity and the symmetry of the order parameters. {\cite%
{Cruz08,Hidenori09,Pratt09,Hirsch2016}}
The most recent findings are of iron-based superconductors (FeBS) having critical temperatures up to 100 K{%
\cite{Feng2015}}. The important issue of the pairing mechanisms and the symmetry of the order parameter in
these materials is still a matter of an extensive debate. 
They, as shown by DFT calculations and confirmed by ARPES, 
are in-fact multiband materials with, either four or five, quasi-2D disconnected Fermi
pockets. {\cite{Singh2008,Kontani2015}} The hole pockets are centred at $\Gamma =(0,0)$ and the electron pockets are centred
at M=($\pi $,$\pi $). The nesting between the electron and hole pockets on the one hand leads to strong spin fluctuations, which favor \(s_\pm\) superconductivity, with the order parameter having the opposite sign for the  electron and the hole pockets.{\cite{Johannes2008,McElroy2003, Mazin2010,Hirschfeld2011,Chubukov2008}} On the other hand it may enhance orbital fluctuations, favoring s++ superconductivity\cite{Hanaguri2010}, with the order parameter, having the same sign for the  electron and the hole pockets. Therefore, such a sign
change of the order parameter between the electron and hole pockets should hint at the possible pairing mechanism. {\cite{Korshunov2011,Monthoux 2001,Golubov2008,Kuroki8, Kordyuk12,Werner2012,Chen2014}} 

Even though
the symmetry of the order parameter was determined for some of the
representative of FeBS, e.g. in the inelastic neutron scattering
experiments, it still does not give the complete picture for all compounds.
The underlying reason is the multiband character of the Fermi surfaces in the FeBS. %
 In this case the order parameter may change sign due to
impurities; as it was demonstrated theoretically
\cite{Efremov2011,Efremov13,Korshunov2014} and experimentally
\cite{shilling2016}  with doping either to d-wave symmetry
\cite{Reid2012,Hafeiz2013,Grinenko2014,Grinenko20142} or change a sign
\cite{Wang2016}. Therefore, a universal tool to ascertain the pairing
symmetry is much needed. In contrast to high $T_{c}$ cuprates, phase-sensitive experiments using
FeBS-based Josephson junctions have not been performed yet. %
The main difficulty for such a multiband superconductor is
the need to design an experimental geometry in such an ingenious way, such
that, the current through one contact is dominated by carriers having
positive sign of order parameter and in the other contact the opposite case
occurs. The isotropic nature of the s-wave fails the effort in this
direction; however, the extended s-wave nature comes directly under the
realm of such experimental investigation.{\cite%
{Golubov2013,Mazin2009,Burmistrova2015,Golubov2009}}

One of the methods for resolving the symmetry of the order parameter, is the study
of the local density of states (LDOS) modulations due to the quasiparticle
interference (QPI), in the presence of impurities; which, could provide
interesting information on the pairing symmetry of the gap function. 
The STM studies of conductance modulations have been
utilized in earlier investigations as the direct probes of the quantum
interference of electronic eigenstates in metals{\cite{Crommie1993}},
semiconductors{\cite{Kanisawa2001}} and cuprates{\cite%
{Hoffman2002,Hanaguri2007,Howald2003}}. In Fe-based superconductors,
theoretical predictions for the dispersion of the QPI vector peaks have been
made with models with electron and hole pockets for the case of $s_{\pm }$%
superconducting order.{\cite%
{Coleman09,Skyora2011,Hirschfeld15,Scalapino2003,Scalapino2012}}

In view of the above discussion, it would be helpful to formulate a model 
for the QPI to reveal qualitative differences between the response in the $s_{\pm}$ and $s_{++}$
pairing states. 
In this work, we formulate such a model for multiband superconductors by
employing the Eliashberg formalism which naturally takes into account the temperature and retardation effects. 
We discuss the temperature dependence of the QPI spectral function and emphasize upon the finite temperature effect on the distinction between the two symmetry cases viz. $s_{\pm}$ and $s_{++}$.
We show, both analytically and numerically, that within the
Born approximation, the quasiparticle interference response function given
as the function of energy has three singularities. Two of these correspond
to the values of the energy gaps and the third depends on
both the gaps and the transfer momentum. We argue that only the lowest
value in the energy singularity may be used as an universal tool for the
determination of the phase shift of the order parameter between the bands. 
We identify the robustness of the sign of response
function peak near the smaller gap value in both the symmetry cases is a promising feature that can be used to identify a pairing symmetry.

The paper is organized as follows. In section II we shortly introduce the main object of the present study, namely the QPI response function and the Eliashberg approach for the single-particle correlation
functions in multiband systems with strong coupling interaction. The
theoretical background to obtain the LDOS and the response function is explained in the section III. where, we
numerically analyse the response function in strong coupling for inter- and
intra-band case. In section IV, the general case of away from ideal nesting
condition with non-zero band ellipticity $\epsilon$ and the shifted Fermi
surface energy $\delta\mu$ is discussed. We show the dependence of QPI
response function on the inherently present large momentum transfer process
that could probe the sign-changing gap symmetry. 
In section V we conclude the paper with the summary of our results.

\section{The Eliashberg Approach}

To find the single-particle correlation functions in multiband systems with
strong coupling interaction we employ the Eliashberg approach {\cite%
{Carbotte90,Scalapino66,Allen82,Marsiglio2008,Scalapino69,Maksimov82,Parker08,Vonsovskii82}%
}. For the sake of simplicity, the consideration here is restricted by assuming the two
bands scenario. The generalization for higher number of bands is straightforward.
Since, the superconducting gap functions have weak momentum dependence, the
systems like Fe-based superconductors can be successfully described in the
frame of quasi-classical Green functions $\hat{\mathbf{g}}_{\alpha}(\omega)$: 
\begin{equation}
\hat{\mathbf{g}}_{\alpha}(\omega)=N_{\alpha}(0)\int d\xi\hat{\mathbf{G}}%
_{\alpha}(\mathbf{k},\omega)
\end{equation}
where $\alpha=a,b$ is the band index and and \(N_\alpha(0)\) is the density of states. In the following, we will use retarded
Green function throughout and therefore we shall omit the index R. In the
Nambu notations the full Green functions have the form: 
\begin{equation}
\hat{\mathbf{G}}_{\alpha}(\mathbf{k},\omega)=\frac{\tilde{\omega}_{\alpha}%
\hat{\tau}_{0}+ \xi_{\alpha, \mathbf{k}}\hat{\tau}_{3}+\tilde{\phi}_{\alpha}%
\hat{\tau}_{1}}{\tilde{\omega}_{\alpha}^{2}- \xi_{\alpha, \mathbf{k}}^{2}-%
\tilde{\phi}_{\alpha}^{2}}
\end{equation}
where, the $\hat{\tau}_{i}$ denote Pauli matrices in Nambu space. Here, $%
\xi_{\alpha,\mathbf{k}}= \epsilon_{\alpha,\mathbf{k}} - \epsilon_F$ is the
dispersion at the Fermi energy. The order parameter $\tilde{\phi}%
_{\alpha}=\tilde{\phi}_{\alpha}(\omega)$ and the renormalized frequency $%
\tilde{\omega}_{\alpha}=\tilde{\omega}_{\alpha}(\omega)$ are complex
functions of the $\omega$. Correspondingly, the quasi-classical $\xi$%
-integrated Green functions can be written:

\begin{eqnarray}
g_{0\alpha}(\omega)&=&-i\pi N_{\alpha}\frac{\omega}{\sqrt{\omega^{2}-\tilde{%
\Delta}_{\alpha}^{2}(\omega)}}, \\
g_{1\alpha}(\omega)&=&-i\pi N_{\alpha}\frac{\tilde{\Delta}_{\alpha}(\omega)}{%
\sqrt{\omega^{2}- \tilde{\Delta}_{\alpha}^{2}(\omega)}},
\end{eqnarray}
where, $\tilde{\Delta}_{\alpha}(\omega) =
\tilde\phi_\alpha(\omega)/Z_\alpha(\omega)$ and $Z_\alpha(\omega) =
\tilde\omega_\alpha(\omega)/\omega$ and are complex functions. The
quasi classical Green functions are obtained by numerical solution of the
Eliashberg equations {\cite{
Scalapino69,Parker08,Maksimov82,Vonsovskii82}}: 
\begin{equation}
\tilde{\omega}_{\alpha}(\omega)\!\!-\!\omega\!=\!\!\!\sum_{\beta}\,\int%
\limits _{-\infty}^{\infty}\!\!\!dzK_{\alpha\beta}^{\tilde{\omega}%
}(z,\omega)Re\frac{\tilde{\omega}_{\beta}(z)}{\sqrt{\tilde{\omega}%
_{\beta}^{2}(z)-\tilde{\phi}_{\beta}^{2}(z)}},  \nonumber
\end{equation}

\begin{equation}
\tilde \phi_\alpha(\omega) \!\! = \sum_{\beta }
\int\limits_{-\infty}^{\infty} dz K^{\tilde\phi}_{\alpha \beta} (z,\omega)
Re \frac{\tilde\phi_\beta(z) }{\sqrt{ \tilde\omega^2_\beta(z) -
\tilde\phi^2_\beta(z) }} .  \label{eq.Elias.2}
\end{equation}
The kernels $K^{\tilde\phi, \tilde\omega}_{\alpha \beta}(z,\omega)$ of the
fermion-boson interaction have the standard form \cite{Scalapino69}: 
\begin{equation}
K_{\alpha \beta}^{\tilde\phi, \tilde\omega} (z, \omega) \!\! = \!\!\!
\int\limits_{-\infty}^\infty \!\! d \Omega \frac{\lambda_{\alpha
\beta}^{\tilde\phi, \tilde\omega} B(\Omega)}{2} \left[ \frac{\tanh \frac{z}{%
2T} + \coth \frac{\Omega}{2T}}{z+ \Omega-\omega - i \delta} \right].
\end{equation}
For simplicity, we use the same normalized spectral function of
electron-boson interaction $B(\Omega)$ obtained for spin fluctuations in
inelastic neutron scattering experiments \cite{Inosov09} for all the
channels. The maximum of the spectra is $\Omega_{sf} = 144~cm^{-1}$, which
determines the natural energy scale \cite{Efremov13}. This spectrum gives a
rather good description of thermodynamical{\cite{Popo2010}} and optical {%
\cite{Charnukha2010,Charnukha2014}}properties in the SC as well as normal
states\cite{Dolgov2011}. 
Moreover, we will use all temperatures and energies, expressed below, in the units of inverse $cm$ (i.e. $cm^{-1}$).  The matrix elements $\lambda^{\tilde\phi}_{\alpha\beta}$ are positive for attractive interactions and negative for repulsive ones. The symmetry of the order parameter in the clean case is determined solely by the off-diagonal matrix elements. The case  $sign \lambda^{\tilde\phi}_{\alpha\beta} =sign \lambda^{\tilde\phi}_{\beta\alpha} >0$ corresponds to $\spp$ superconductivity and $sign\lambda^{\tilde\phi}_{\alpha\beta} =sign \lambda^{\tilde\phi}_{\beta\alpha} <0$ to $\spm$ case.  The matrix elements
$\lambda^{\tilde\omega}_{\alpha\beta} $ have to be positive and  are chosen $\lambda^{\tilde\omega}_{\alpha\beta} = |\lambda^{\tilde\phi}_{\alpha\beta}|$. Further for simplicity we will omit the subscripts $\tilde \omega$ and $\tilde \phi$ denoting $\lambda^{\tilde\phi}_{\alpha\beta} = \lambda_{\alpha\beta}$ and $\lambda^{\tilde\omega}_{\alpha\beta} = |\lambda_{\alpha\beta}|$. Additionally, we would also use notation \(\Delta_a\) and \(\Delta_b\) for the real band gap energy values.

%We start with the four-band model consisting from two circular hole bands
%around $\Gamma=(0,0)$ point and two elliptical electron bands around $M_1 =
%(\pi,0)$, $M_2 = (0,\pi)$ points. Since, the bands appear in Eq.(\ref%
%{eq.qpi.general}) pairwise, one can restrict the consideration of the
%four-band model to a two-band model by considering the pairs i.e. electron-
%electron, hole-hole and electron-hole bands, designated as bands $a,b$ i.e.
%the hole-like and electron-like effective bands, respectively. 

\begin{figure}[]
\includegraphics[scale=0.45]{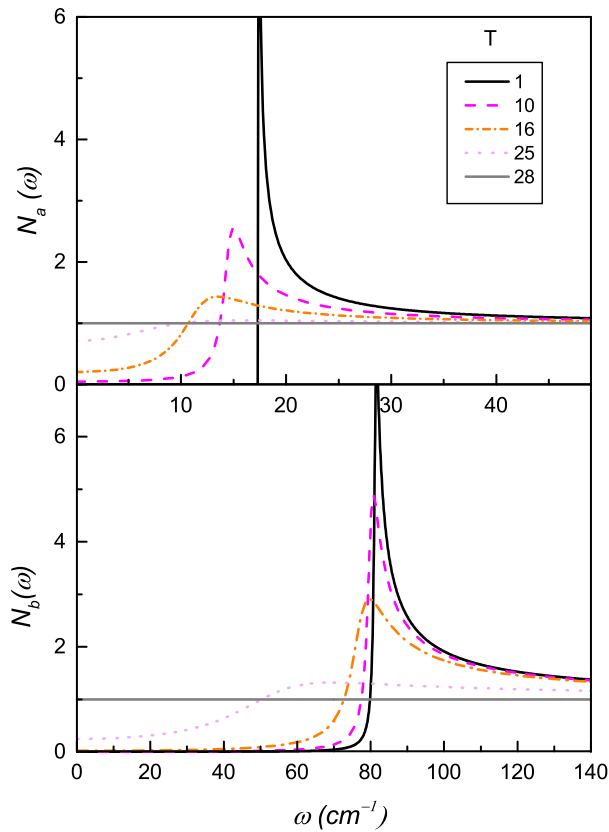}
\caption{Density of states for the bands $a$ and $b$ calculated in strong
coupling at various temperatures. The coupling constants are $\protect\lambda%
_{aa}=0.5$, $\protect\lambda_{ab}=0.2$, $\protect\lambda_{ba}=0.1$, $\protect%
\lambda_{bb}=3$. The superconducting critical temperature is $T_c=28 \, cm^{-1}$. The
DOS is normalized with respect to the normal state and is set equal to 1 for each band. }
\label{fig.DOS}
\end{figure}
In the strong coupling approach, as opposed to the weak coupling limit, the
gap functions are complex and frequency dependent $\tilde\phi_\alpha=\tilde%
\phi_\alpha(\omega)$. One of the consequences is the broadening of the
quasiparticle peaks and appearing of the finite density of states $%
N_\alpha(\omega) = - \frac{1}{\pi}Im g_{0 \alpha}(\omega)$ at zero energy.
This behavior is illustrated in Fig.\ref{fig.DOS}. At zero temperature,
DOS in the strong coupling approach exhibits the coherence peak $N(\omega)
\propto 1/\sqrt{\omega-\Delta}$ for $\omega\geq\Delta(\omega)$ and zero for $%
\omega<\Delta(\omega)$ quite similar to the weak coupling case. But at
finite temperatures, DOS becomes finite for $\omega<\Delta(\omega)$ and the
coherence peak is smeared out. This behavior is completely different from
the weak coupling approximation. The reason is that the gap function $%
\Delta(\omega)$ in strong coupling approximation is a complex function.
Accounting for the frequency dependence of the gap functions on the QPI is
the key issue of the present work. At the same time, one has to point out
that the DOS measurements are unable to distinguish between $s_{++}$ and $%
s_{\pm}$ order parameter symmetries (as is seen from Eq.3, DOS depends on $%
|\Delta(\omega,T)|$. A phase-sensitive QPI calculation is needed to bring
out the contrast between the two types of pairing symmetries.

\section{Quasiparticle interference.}

The STM measures the differential conductance; which, is proportional to the
local single particle density of states $N (\mathbf{r}, \omega)$: 
\[
\frac{d I }{d V}(\mathbf{r},\omega) \propto |M(\mathbf{r})|^2 N (\mathbf{r},
\omega),
\]
where, $M(\mathbf{r})$ is the local tunnelling matrix element. The local
density of states is related to the single particle retarded Green functions 
$G^R(\mathbf{r},\mathbf{r},\omega)$:

\begin{equation}
N(\mathbf{r}, \omega) = - \frac{1}{\pi} \mbox{Im} Tr\left[ \frac{1+\tau_3}{2}
\hat G^R(\mathbf{r},\mathbf{r},\omega ) \right]
\end{equation}
Here, $Tr[..]$ is taken over both Nambu and band indices. Although the
tunnelling matrix element may be important in the multiband case, sharpening
the spectral weight contribution of some orbitals, the strong coupling does
not affect the tunnelling matrix element. Since, we want to focus here on the
effects of strong coupling the consideration is restricted by the impact of
a single impurity on the local density of states. In the linear response
approximation the perturbation of the density of states form due to an
impurity with the point-like scattering $\displaystyle\hat{U}(\mathbf{r}%
)=U_{\alpha \beta}\delta(\mathbf{r}) \tau_{3}$ reads \cite{Dolgov2009}: 
\begin{widetext}
\ba
\delta N(\mathbf{r},\omega) &=& -\frac{1}{\pi}Im\sum_{\alpha, \beta} Tr\left[\frac{1+\tau_{3}}{2}\int d V ^{\prime\prime}\hat{G}_{clean}^{\alpha}(\mathbf{r}-\mathbf{r}^{\prime\prime},\omega)\hat{U}_{\alpha \beta}(\mathbf{r}^{\prime\prime})
\hat{G}_{clean}^{\beta}(\mathbf{r}^{\prime\prime}-\mathbf{r},\omega)\right]
\label{eq.dN}
\ea
\end{widetext}
for $\omega >0$. The negative values of $\omega$ can be obtained by
substitution $\tau_3 \to - \tau_3$. Since, in the response function, the bands are considered pairwise within the Born approximation; we will consider below the scattering between two bands, having in mind that one has to sum up the full response function afterwards.  Considering Eq.(\ref{eq.dN}) in the
momentum space and keeping only the interband impurity scattering, which
gives the leading contribution for the momentum $\mathbf{q}$ close to the
interband vector $\mathbf{Q}$, we define the QPI response function $I(%
\mathbf{q},\omega)$ as: 
\[
\delta N(\mathbf{r},\omega) = U_{a b} \int \frac{d^2 q}{(2\pi)^2} e^{i 
\mathbf{q r}} I(\mathbf{q}, \omega) 
\]
The response function is given by the following expression: 
\begin{widetext}
\be
I(\mathbf{q},\omega)=-\frac{1}{2 \pi} \int \frac{d^2 p}{(2\pi)^2} Im Tr\left[\tau_{3} \hat{G}_{clean}^{a}(\mathbf{q+p},\omega)\tau_3
\hat{G}_{clean}^{b}(\mathbf{p},\omega)\right] + (a \leftrightarrow b).
\label{eq.qpi.general}
\ee
\end{widetext}

\subsection{The model.}

We apply the above formulation to develop the model for the general pnictide
case as discussed below. In the low energy limit considered here, the
spectrum near to the Fermi-level can be linearized: 
\begin{equation}  \label{eq:11}
\xi_b(\mathbf{p}+\mathbf{q}) \approx \beta \xi_a(\mathbf{p}) + \epsilon \cos
2 \theta +\delta \mu.
\end{equation}
Here,  $sign \beta>0$   for impurity scattering between two
electron  or  two hole  bands, while $sign \beta<0$  for scattering between 
electron and hole bands. We assume  constant density of states
$\displaystyle N_{\alpha}=\int\delta(\xi_{\alpha,\mathbf{p}}) d^{2}p
/(2\pi)^{2}$ and $|\beta| = v_b/v_a$. Where, the $v_{a,b}$ are the Fermi
velocities for the two bands. The parameter $\epsilon = \mathbf{k}_{Fy}
\mathbf{v}_b- \mathbf{k}_{F x} \mathbf{v}_b$  characterizes the ellipticity
of the electron bands; where, $\mathbf{k}_{F y}\, \text{and} \,
\mathbf{k}_{Fx}$ are the electron band Fermi wave vectors. Here,
$\theta$ is angle between the vector \(\mathbf{p}\)
and \(\mathbf{q}\). We have $\epsilon =0$ for scattering between two hole bands; otherwise
$\epsilon$ is finite. Finally, $\delta \mu$ accounts the relative energy shift
of the Fermi-surfaces and is given by the relation \(\delta\mu
=(\mathbf{k}_F \mathbf{v}_F)_a-(\mathbf{k}_F \mathbf{v}_F)_b\).

\subsection{Scattering at $q=Q$.}

The direct integration over $\xi$ and the angle gives the following
expression 
\begin{equation}
I\! ( \mathbf{q} =\mathbf{Q} ,\omega)\! = -\frac{\sqrt{N_{a} N_b}}{2} Im[
K(\omega) F(\omega)],  \label{eq.qpi.Q}
\end{equation}

where, the coherence factor $K(\omega)$ is 
\begin{equation}  \label{eq:13}
K(\omega)=\left[ \frac{\tilde\Delta_{a}\tilde\Delta_b -\omega^2 }{E_a E_b}%
\pm 1\right]
\end{equation}
and 
\begin{eqnarray}  \label{eq:14}
\!\!\!&F&\!\!\!(\omega)= \frac{1}{\sqrt{|\beta|^{-1} \epsilon^2 - \! \left(\!%
\sqrt{|\beta|}\!Z_{a}\! E_a+\sqrt{|\beta|^{-1}} (\!Z_{b}\!E_b +
\delta\mu) \right)^{2}}}\!  \nonumber \\
&+&\!\!\frac{1}{\sqrt{|\beta|^{-1} \epsilon^2-\! \left(\!\sqrt{|\beta|}%
\!Z_{a}\!E_a +\sqrt{|\beta|^{-1}}(\!Z_{b}\!E_b - \delta
\mu)\right)^{2}}}\!.  \label{eq.F1}
\end{eqnarray}
Here, $E_\alpha = \sqrt{\omega^{2}-\tilde\Delta_{\alpha}^{2}}$ is the
quasiparticle energy spectrum. In the coherence factor $K(\omega)$ the sign
"+" corresponds to the scattering between two electron or two hole bands,
while "-" to the case of the scattering between electron and hole bands. One
can immediately notice that the response function for intraband scattering
at $q=0$ vanishes due to the coherence factor for all $\omega$. In our
study, we have focussed completely on the inter-band interactions aspect of
the phenomenon. This implies the choice of the "-" sign in the relation for
the coherence factor given by Eq.(\ref{eq:13}).

\subsubsection{Zero ellipticity}

The hole bands around $\Gamma$ point can be considered in a good
approximations as circle ($\epsilon=0$).  For simplicity, in discussing the two
cases for the band ellipticity \(\epsilon\), we shall assume the system to
be in the weak coupling regime; and hence, take
\(\tilde{\Delta}_{\alpha/\beta}\) to be real and write it as
\(\Delta_{a/b}\) for the smaller (hole band) and larger (electron band) band
gap energy, respectively. We start with perfectly matching hole bands ($%
\delta\mu=0$), having the gap functions $\Delta_a(\omega)>\Delta_b(\omega)$.
The same ratio of the gap functions is used in the relation below. For the sake of simplicity, we put \(\beta =1\) for further analysis.  The
function $I(\omega)$ diverges as $\displaystyle{\pm Re[1/\sqrt{%
\omega-\Delta_b}]}$ for $\omega>\Delta_b$ and as $1/\sqrt{%
|\omega-\Delta_a|}$ for $\omega$ close to $\Delta_a$. The sign in front of
the first singularity depends on the symmetry of the order parameter. Sign "$%
-$" corresponds to $s_\pm$ superconductivity, while "$+$" for $s_{++}$
superconductivity. However, the sign in front of the second singularity does
not depend on the superconducting order parameter symmetry. The mismatch of
the bands creates non-zero $\delta \mu$, which, considerably changes the $%
\omega$-dependence of the response function. For very large values of $%
\delta\mu$, there is an additional dip at $\omega^* = \sqrt{
(\Delta_a^2 + \Delta_b^2 +\delta \mu^2 )^2- 4\Delta_a^2 \Delta_b^2}%
/(2|\delta\mu|)$ at energy greater than $\Delta_b$. The divergence for
energies near to $\Delta_b$ remains as $1/\sqrt{\omega-\Delta_a}$ for 
$\omega^*>\Delta_a$. The case for finite band ellipticity is considered
below.

\subsubsection{Finite ellipticity}

For scattering between two electron bands, the essential role is played by
the ellipticity of the electron bands i.e. $\epsilon$. Here, we have
distinct cases: a) $|\epsilon| + |\delta\mu|<\Delta_b$, b) $|\epsilon| +
|\delta\mu|>\Delta_b$ and $||\epsilon| -|\delta\mu||<\Delta_a$, c) $%
||\epsilon| -|\delta\mu||>\Delta_a$. For the case a) one finds the behaviour
similar to the scattering between two hole bands i.e. the appearance of a
dip. In the case b) in addition to $1/\sqrt{\omega-\Delta_b}$ and $1/\sqrt{%
\Delta_a-\omega}$ a new divergence of $1/\sqrt{\omega-\omega_1}$ appears at $%
\omega_1 = \sqrt{ (\Delta_a^2 + \Delta_b^2 +(\delta \mu+|\epsilon|)^2)^2-
4\Delta_a^2 \Delta_b^2}/(2(|\delta\mu|+|\epsilon|))$. In the case c) one
additional divergence $1/\sqrt{\omega -\omega_2}$ occurs at $\omega_2 = 
\sqrt{ (\Delta_a^2 + \Delta_b^2 +(\delta \mu-|\epsilon|)^2)^2- 4\Delta_a^2
\Delta_b^2}/(2||\delta\mu|-|\epsilon||)$.

\subsection{Scattering at $q= Q + \tilde q$}

Now, we consider the quasiparticle interference due to interband scattering
at the vector $\mathbf{\tilde q} = \mathbf{q}-\mathbf{Q}$. For small $\tilde
q$ one can use the approximation $\xi_b(\mathbf{p}+\mathbf{q})  \approx \beta \xi_a(%
\mathbf{p})+ \epsilon \cos 2 \theta + v_b \tilde q \cos(\theta -
\phi)+\delta \mu $, where $\phi$ is the angle between the vector $\tilde{\mathbf{q}}$
and $\mathbf{Q}$. The F-function in Eq.(\ref{eq.qpi.Q}) has the form: 
\begin{widetext}
\begin{eqnarray}\label{eq:15}
F(\omega,\phi) = \left\langle
 \frac{\!\sqrt{|\beta|} Z_{a} E_a +\sqrt{|\beta|^{-1}}Z_{b} E_b }{(  \!\sqrt{|\beta|}Z_{a} E_a \!+\!\sqrt{|\beta|^{-1}} Z_{b} E_b )^2\!+ |\beta|^{-1}\left(\epsilon \cos(2 \theta)\! + \! v_b \tilde q \cos (\theta - \phi) \!+\!  \delta \mu \right)^2} \right\rangle_{\!\!\theta\,},
 \label{eq.F2}
\end{eqnarray}
\end{widetext}
where $\langle ...\rangle_\theta$ is the averaging over the angle. The
integration over the angle can be easily performed in two limits of $%
\epsilon\gg \mathbf{v}_{b}\mathbf{\tilde{q}}$ (setting $\mathbf{v}_{b}%
\mathbf{\tilde{q}}=0$) and $\epsilon\ll \mathbf{v}_{b}\mathbf{\tilde{q}} $
(setting $\epsilon =0$). In the second limit we recover expression similar
to Eq.(\ref{eq.F1}) with substitution $\epsilon \to \mathbf{v}_{b}\mathbf{%
\tilde{q}}$.

\section{Numerical Analysis and Results}

In the following, we will apply the above general formulation to Fe-BS, using the electron-boson spectral function, successfully used by Popovich
et. al.\cite{Popo2010} for the thermal studies and by Charnukha et. al.\cite{Charnukha2010} for optical conductivities for the description of BaKFeAs at optimal doping. 
According to \cite{Charnukha2010}, the original four-band model for \(Ba_{1-x}K_x Fe_2 As_2\) can be reduced to an effective two-band model,
where the first band is formed by the inner hole pocket with the gap \(\Delta_a\),
while the second band with the gap \(\Delta_b > \Delta_a\) consists from combination of two electron pockets and outer hole pocket.
Within this two-band model we will calculate the response \(I(\mathbf{q},\omega)\) at  \(\mathbf{q}\) values around the nesting vector \(\mathbf{Q} = (\pi, \pi)\).

\begin{figure}
	\includegraphics[scale=0.45]{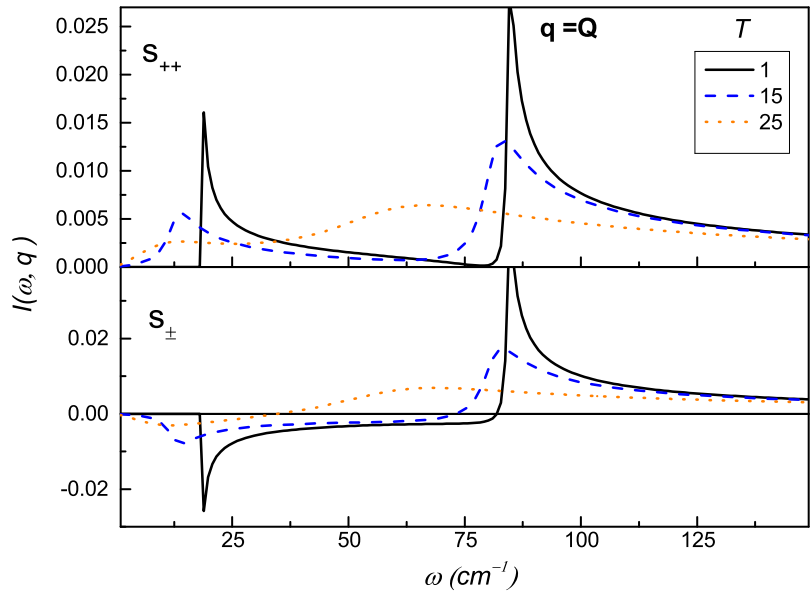}
	\caption{The response function I($\protect\omega$) for the $%
		s_{++}\, \text{and}\, s_{\pm}$ case with the strong coupling $%
		\protect\lambda$-matrix defined as ($\protect\lambda_{aa}$=3, $\protect%
		\lambda_{ab}$=${\pm}$0.2, $\protect\lambda_{ba}$=${\pm}$0.1, $\protect\lambda%
		_{bb}$=0.5) and \(T_c\) =28 \(cm^{-1}\).}
	\label{Fig.strongsppm}
\end{figure}

The model is studied in the
beginning with $\epsilon=\delta\mu$ =0 (non-FeBS case), and later in the paper, we
would consider finite values of $\delta\mu$ and $\epsilon$, as is the case with
pnictides. Hence, the model has broader implications to other high $T_c$
superconductors. In this case, we have only two characteristic energy values, namely the energies of the gaps \(\Delta_a\) and \(\Delta_b\). Our purpose is to identify certain peculiarities of the QPI response for the 
$s_{++}$ and $s_{\pm}$ pairing symmetries. The resulting real-valued energy gaps in $N_\alpha(\omega)$, as
discussed in Fig.1, are $\Delta_{a}$ = 17 \(cm^{-1}\), while $\Delta_{b}$
= 83 \(cm^{-1}\) at T = 0, which gives a gap ratio $\Delta_{b}/
\Delta_{a}$ = 4.82.

In Fig. 2, we discuss the temperature evolution of the response function for $s_{++}$ and $s_{\pm}$ symmetry. 
First, at
temperature T =1, the QPI response vanishes for $\omega < \Delta_a$ for both  $s_{++}$ and $s_{\pm}$ order parameters; since, there is no excitation at the energy below \(\Delta_a\). 
In the whole temperature range, the response function for $s_{++}$ superconductivity is positive for all values of $\omega$; while in the $s_{\pm}$ case, for energies around the smaller gap, it is negative.  As the temperature increases, the response related to the $s_{\pm}$ symmetry turns positive at much lower energies, while for $s_{++}$ case, the response peak shows
 a gradual shift towards the energy interval between the two band gaps. To sum up, the main
feature that help us to distinguish between the response behaviour for the $%
s_{++}$ and $s_{\pm}$ symmetry cases is the robustness of the sign of the peaks near the small band gap $\Delta_{a}$ over a broad range of $T < T_{c}$.

\begin{figure}
	\centering
	\subfloat[\(s_{++}\) symmetry]{{\includegraphics[width=0.45\linewidth]{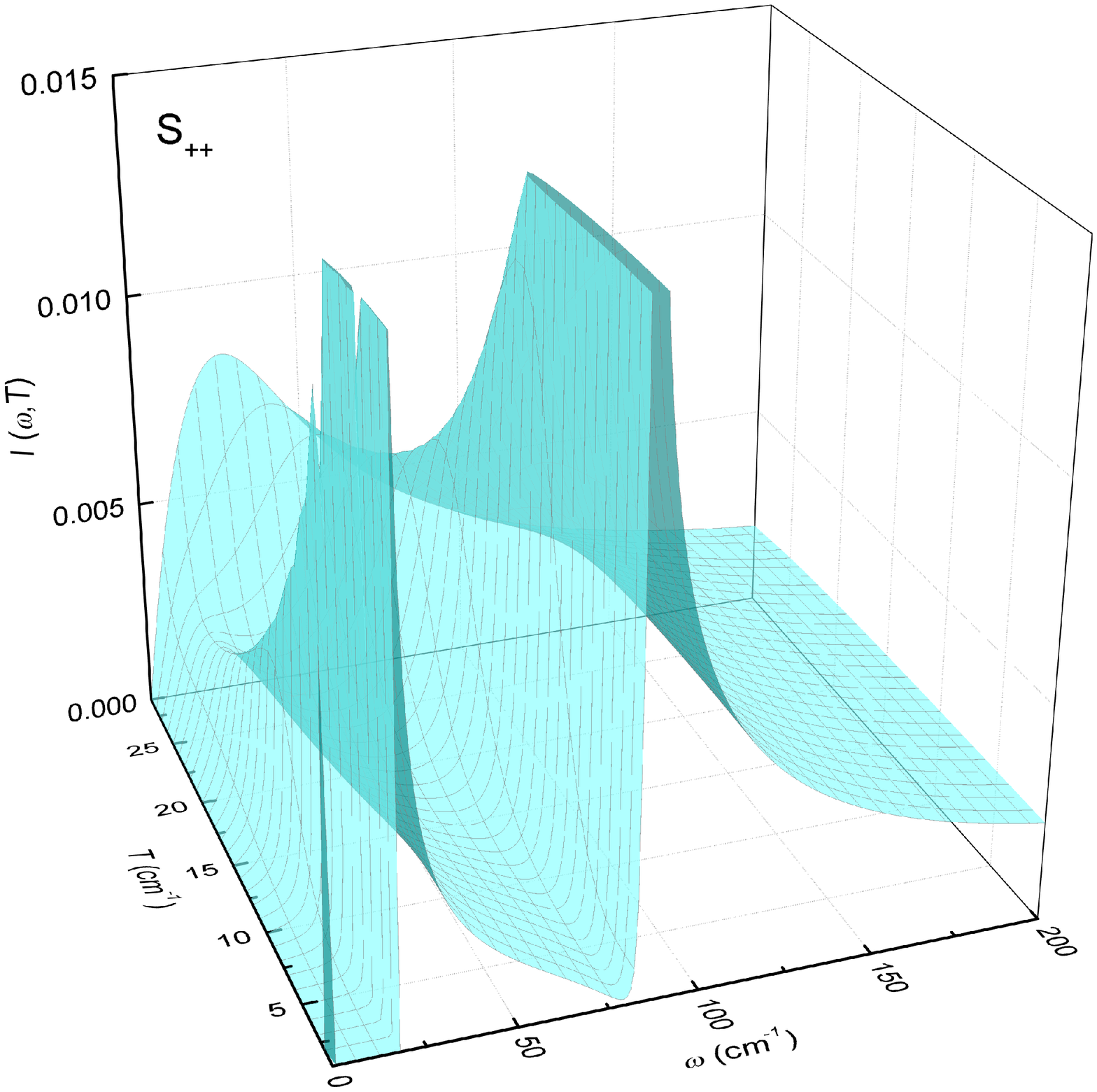} }}
	\qquad
	\subfloat[\(s_{\pm}\) symmetry]{{\includegraphics[width=0.45\linewidth]{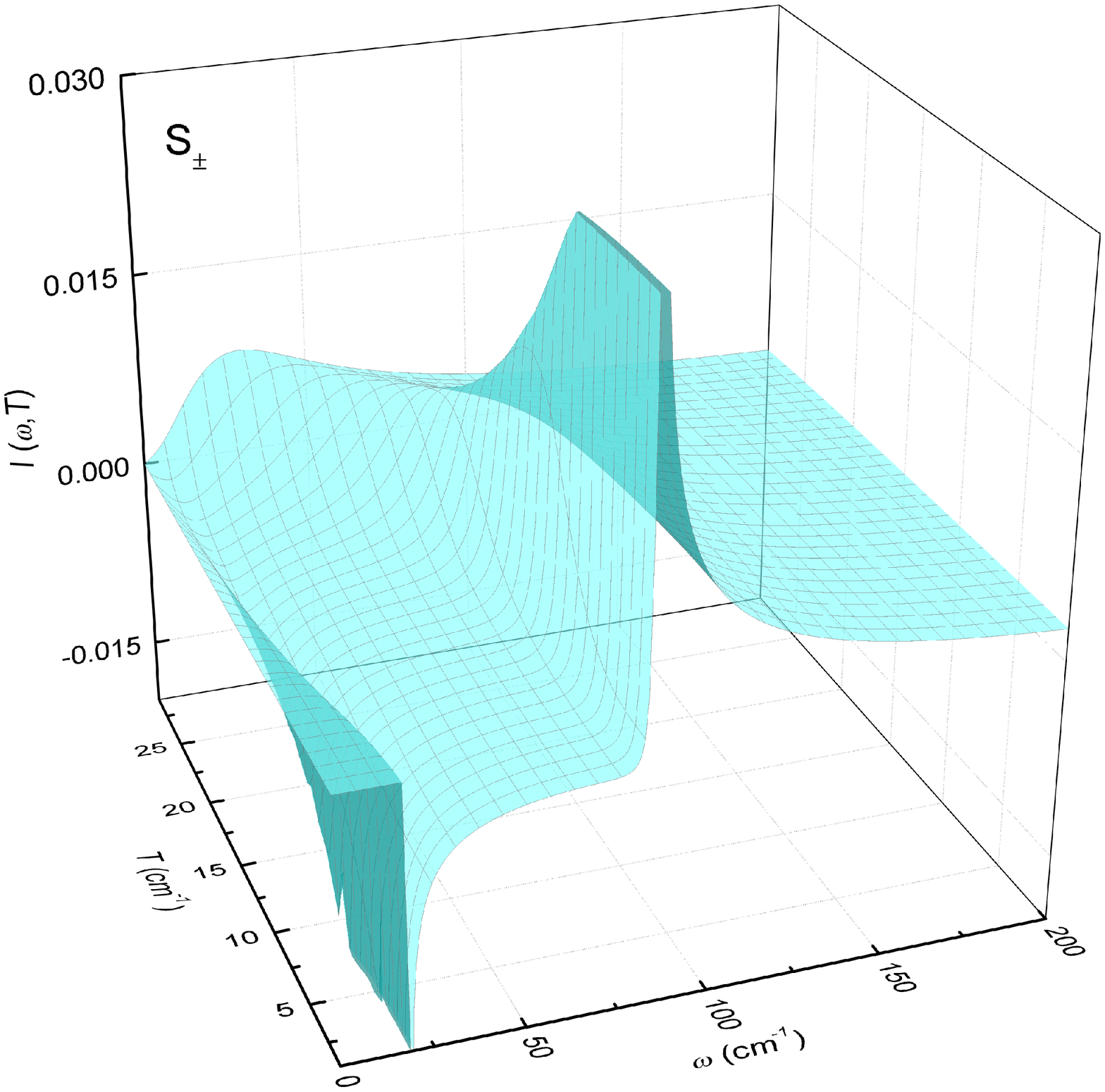} }}
	\caption{3D plot of the response function vs temperature at fixed energy $
		\protect\omega$ for $s_{++}$ (upper) and $s_{\pm}$ (lower) cases,
		respectively. The coupling parameters are the same as used above and the
		transition temperature, $T_c$=28 \(cm^{-1}\).}
	\label{Fig.respvstemp}
\end{figure}

The Fig. 3 represents the 3D plot depicting the variation of $I(\omega)$
simultaneously with temperature T and energy $\omega$ for the case of
perfect nesting i.e. $\mathbf{q} =\mathbf{Q}$. For $s_{++}$ symmetry, at low
temperatures and $\omega \le \Delta_a$, we consider the slice in the
region $0 <T< 10$ that shows a small sharp peak which dips smoothly as the
temperature rises. Moving towards high energies and at low temperatures,the
peak around the second band gap energy is very strong and decays much slower
with rising temperature and energy than compared to the first peak. While in
the $s_{\pm}$ case, we see the difference for the first band peak as the
response at low temperature and low energy is inverted (at $\omega\leq$ 20)
and has large magnitude. This is the main feature that reflects throughout
our analysis. The peaks around the first band gap energy are a robust
indication of the difference between the two symmetry cases viz. $s_{++}$
and $s_{\pm}$.

In the region of sub-gap energies and low temperatures, the $s_{++}$
response shows a negative gradient while the $s_{\pm}$ curve is almost flat and is negative;
and for the same energies at high temperatures, the behaviour is similar for
both the symmetries and hence it is indistinguishable in this region. Beyond
that, the graph shows a monotonically decreasing trend for both $s_{++}$ and 
$s_{\pm}$ response function and does not provide any interesting
distinguishable feature apart from the greater signal strength for $s_{++}$
curve than the latter. As we move to the higher temperatures, a bump in the
response function arises, which is appreciably diffused and broadened as
compared to the ones at low temperatures. This behaviour of response
function is same, in both $s_{++}$ and $s_{\pm}$ case for $T > $25 as stated
for the Fig. 2.

\begin{figure}[]
	\includegraphics[width=0.7\columnwidth]{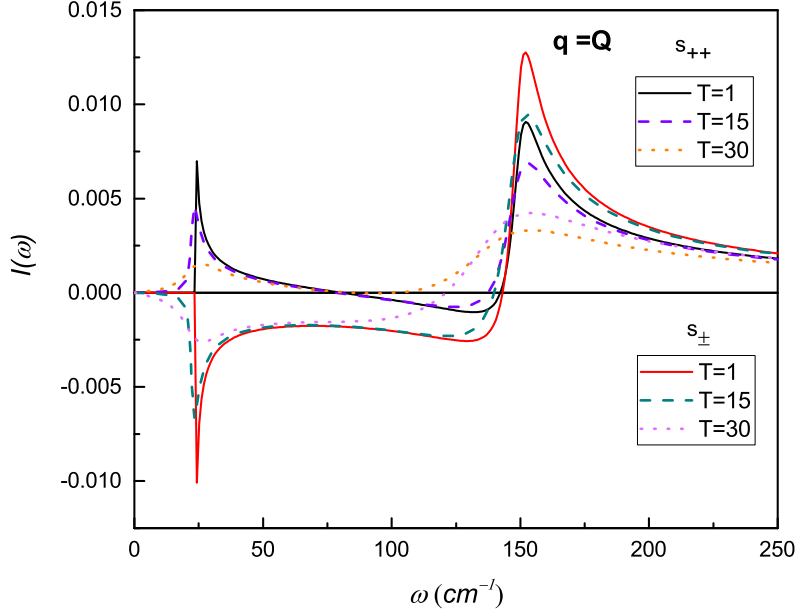}
	\caption{The QPI response function for the $s_{++}$ and $s_{\pm}$ case at
		very strong couplings $\tilde{\protect\lambda}$ i.e. $\protect\lambda_{aa}$%
		=1,$\protect\lambda_{bb}$=6,$\protect\lambda_{ab}$=0.4, $\protect\lambda_{ba}
		$=0.2 with the transition temperature $T_c$ =46 \(cm^{-1}\).}
	\label{Fig.strongLambda}
\end{figure}

For Fig. 4, we have I($\omega$) vs energy $\omega$ plotted at various
temperatures with very strong coupling parameters $\tilde{\lambda}$ and a
raised transition temperature i.e. $T_c$ = 46. In the subgap region, for the s$_{++}$ case,
we identify a peculiar behaviour of the response function (compare Fig. 2)
as it goes to negative values and peaks just like the response for $s_{\pm}$
case. In summary, for the energies near the second band gap,
the behaviour of response function for both the symmetry cases is
indistinguishable apart from their relative strengths. However, we again
observe that the response peaks near the smaller gap are a defining and
distinguishing feature even for a very strong coupling case. 

\begin{figure}[]
\includegraphics[width=0.45\columnwidth]{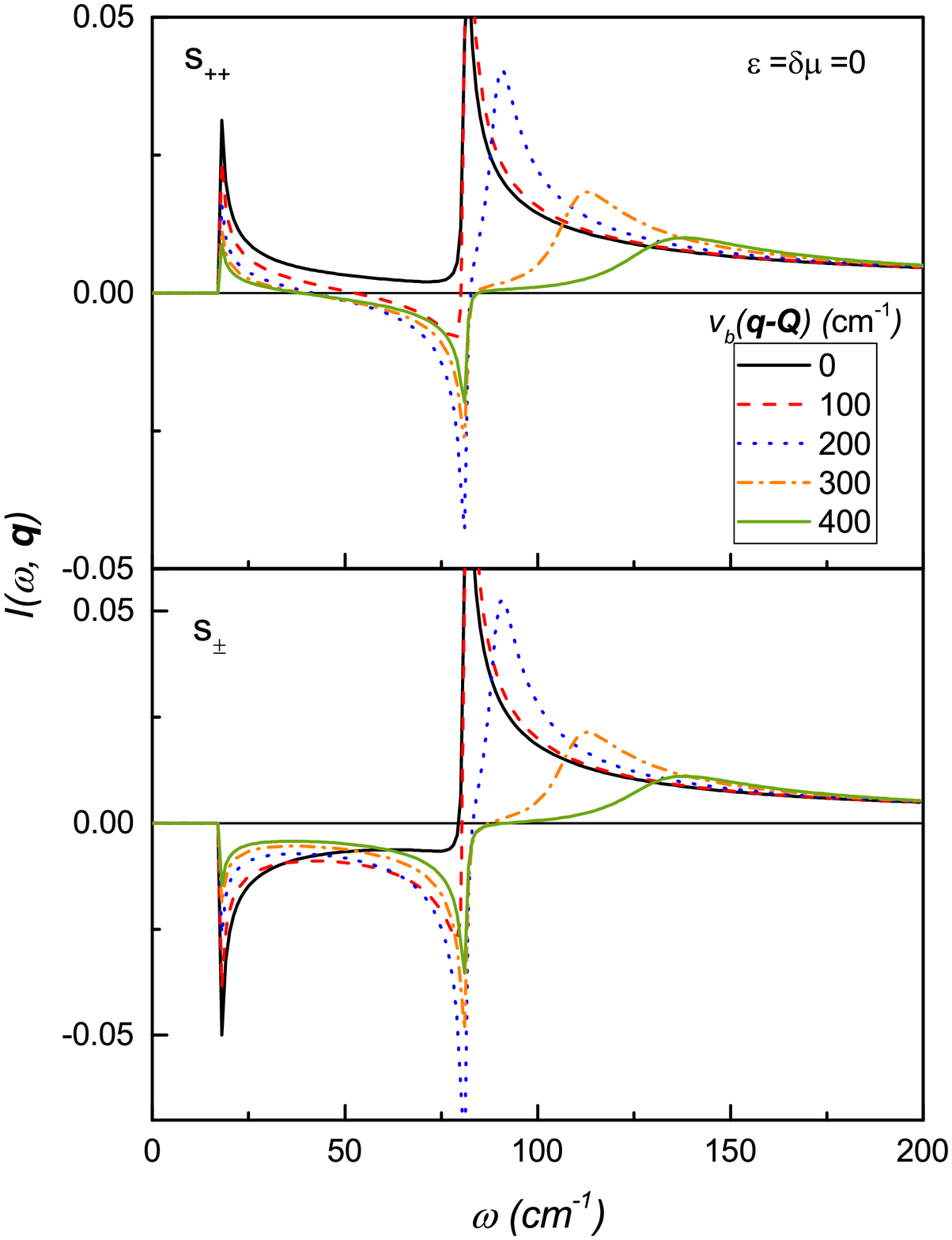}
\caption{The 2D plot of the QPI response function I($\protect\omega$) vs $%
\protect\omega$ and the momentum $\mathbf{q}$ for the
strong coupling case with $\protect\epsilon = \protect%
\delta\protect\mu$ = 0 at temperature T = 1. The values of the coupling constants are $\protect%
\lambda_{aa}=0.5$, $\protect\lambda_{ab}=0.2$, $\protect\lambda_{ba}=0.1$, $%
\protect\lambda_{bb}=3$.}
\label{Fig.vfq2D}
\end{figure}

In the following, we present the study of the response function behaviour
with respect to the changes in parameters such as the ellipticity $\epsilon$
of the electron-like bands, the shifted Fermi energy $\delta\mu$ between the
hole-like and the electron-like bands and the experimentally tunable
electron momentum parameter $v_b \tilde{q}$; which, points in the radial direction to the
electron band Fermi surface. Here, $\mathbf{\tilde{q}}$ is tuned in order to
obtain the correct matching condition for the shifted Fermi energy surface,
as discussed later, and to study the response behaviour closer to the region
of Fermi surface instability, as followed from Eq.(\ref{eq:15}).

In the Figs. 5 and 6, we plot in 2D and 3D, the behavior of $\mathbf{q}$%
-resolved response function for both the symmetry cases, with variation in
the electron-like quasiparticle momentum $\mathbf{\tilde{q}}$ using Eq.(\ref%
{eq:15}) and setting the ellipticity and surface energy to zero. We also
assume that the momentum vector $\tilde{\mathbf{q}}$ is directed along ${\mathbf{Q}}$ and hence, the angle $\phi$ =0. The finite value of $%
\mathbf{\tilde{q}}$ relates to the fact that we are probing the Fermi
surface of the electron-like band pocket. We have $||\epsilon|
- |\delta\mu||< \Delta_a$ satisfied in this case. For the peak near
larger band gap energy, the amplitude and the sign of the peak are robust
and distinguishing features. 

\textbf{\begin{figure}
		\centering
		\subfloat[\(s_{++}\) symmetry]{{\includegraphics[width=0.45\columnwidth]{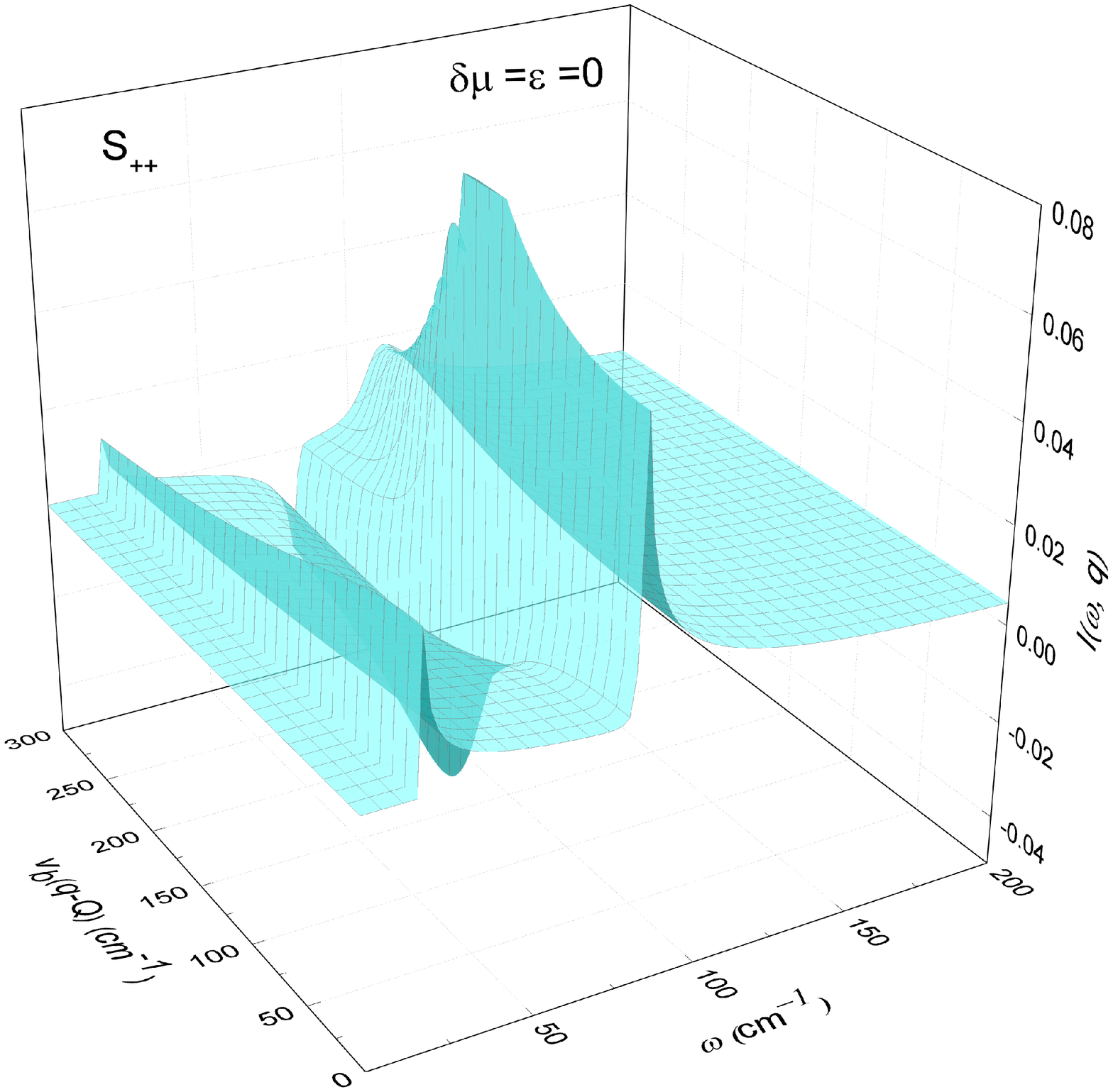} }}
		\qquad
		\subfloat[\(s_{\pm}\) symmetry]{{\includegraphics[width=0.45\columnwidth]{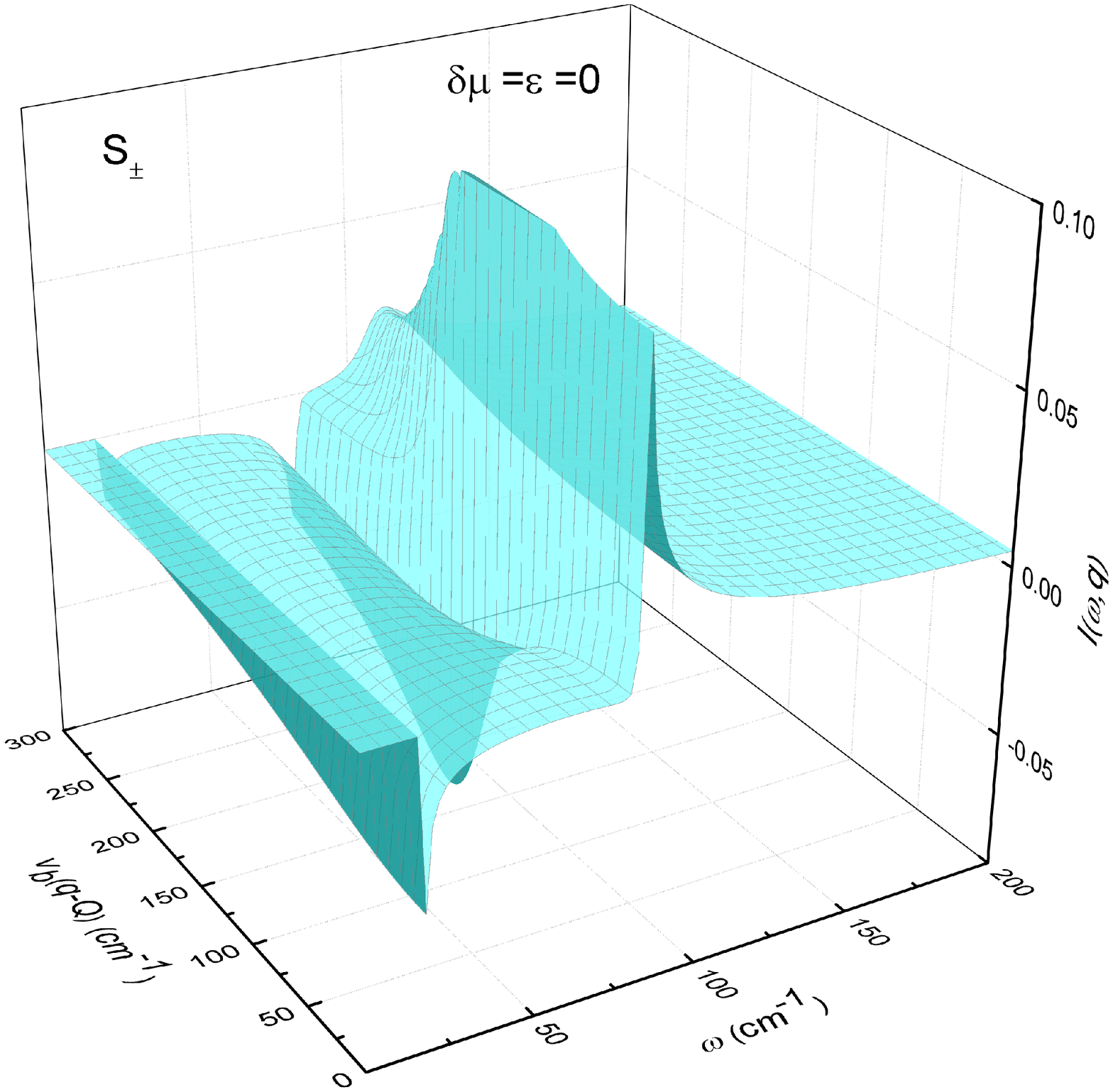} }}
		\caption{The 3D plot of the QPI response function I($\protect\omega$) vs $%
			\protect\omega$ and momentum $\mathbf{q}$ with the zero ellipticity $%
			\protect\epsilon$=0 and zero shifted Fermi surface energy $\protect\delta%
			\protect\mu$=0 for the strong coupling case at temperature T = 1. The values
			of the coupling constants are $\protect\lambda_{aa}=0.5$, $\protect\lambda%
			_{ab}=0.2$, $\protect\lambda_{ba}=0.1$, $\protect\lambda_{bb}=3$..}
		\label{Fig.eps4}
\end{figure}}

We see that the energy dependence of the response function at finite $\mathbf{\tilde{q}}$   shows three peaks. Two of these are momentum independent and correspond to the gaps in the bands \(\Delta_a\) and \(\Delta_b\), while the third peaks has a strong  $\mathbf{\tilde{q}}$ dependence. The strong difference between $s_{++}$ and $s_{\pm}$ symmetries we see only for the first peak at he energy of the small gap. For the $s_{\pm}$ order parameter the response function at $\omega = \Delta_a$ is negative, while for $s_{++}$ it is positive. It leads to the conclusion that for determining the symmetry of the order parameter, one has to consider the response function at momenta close to the nesting vector $\mathbf{Q}$, and find the momentum independent peaks. The smallest of these peak will determine the symmetry of the order  parameter.

The QPI response at energies close to the second gap \( \Delta_b \) is shown in Fig.2 for $\mathbf{\tilde{q}}$ i.e. ($\mathbf{q} - \mathbf{Q}$) has opposite sign compared to the results presented by Hirschfeld et. al\cite{Hirschfeld15}, using a similar
model in the weak-coupling regime. The results presented in  Figs. 5 and 6 clearly demonstrate that with the increase of \(\tilde{\mathbf{q}}\), the sign of the second peak reverses. In this respect, our results do not contradict to those of \cite{Hirschfeld15},  $\mathbf{q}$-integrated response function was presented
to be dominated by large $\mathbf{q}$ values. 
Moreover, our $\mathbf{q}$-resolved results provide more information about the QPI response behaviour. In particular, for non-zero ellipticity $\epsilon$ or the non-zero chemical potential shift $\delta \mu$, we have obtained additional mode at energies above $\Delta_b$ as shown in  Figs. (5-11). 

Hence, we again argue that the peak near the first band gap energy i.e. $%
\omega \approx \Delta_a(\omega)$ is the only strong distinguishing feature
for the phase sensitive experiments for the gap symmetry measurements.

\begin{figure}
\includegraphics[width=0.45\columnwidth]{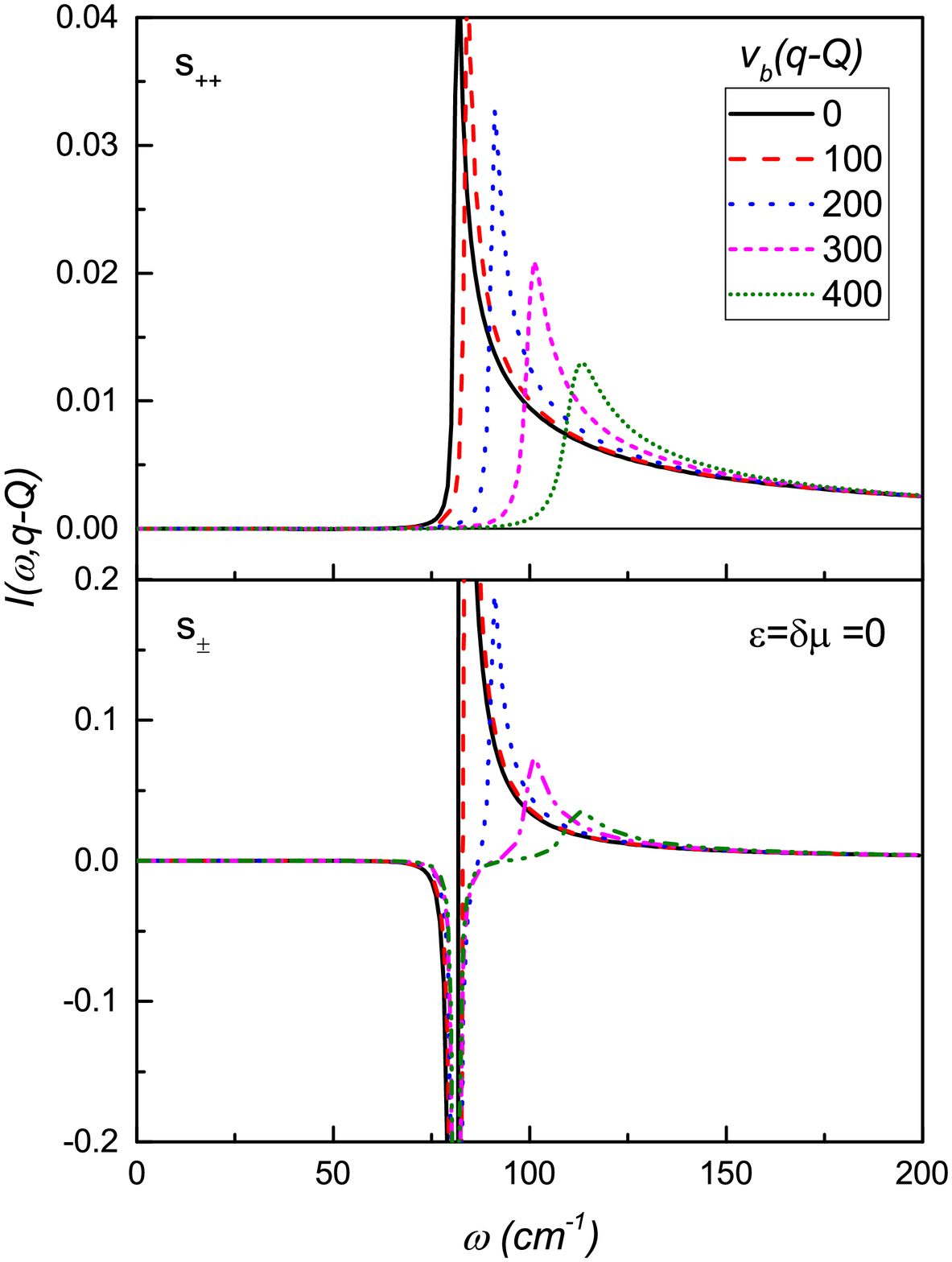}
\caption{The 2D plot of the QPI response function I($\protect\omega$) vs $%
		\protect\omega$ and momentum $\mathbf{q}$ for the
		strong coupling case, with $\Delta_a = \Delta_b $, and $\protect\epsilon = \protect%
		\delta\protect\mu$ = 0 at temperature T = 1; with the coupling constants given as $\protect%
		\lambda_{aa}=0.5$, $\protect\lambda_{ab}=0.2$, $\protect\lambda_{ba}=0.1$, $%
		\protect\lambda_{bb}=3$.}
	\label{Fig.5vfq2D}
\end{figure}

So far, we explored the region around the nesting vector \(\mathbf{Q}=(\pi,\pi)\) with scattering between the smaller/inner hole-like band to the outer/larger averaged electron-like band. Now, we focus on the scattering of the quasiparticles from electron-like band to the outer hole-like band with larger gap value i.e. \(\tilde{\Delta}_{a2}(\omega) \rightarrow \tilde{\Delta}_{b1/b2}(\omega) \). In Fig. 7, we plot the response function for various values of the electron-like quasiparticle  momentum \(\mathbf{\tilde{q}}\) over full spectrum of energy \(\omega\) with equal band gap functions.  For this, we modify Eq.(\ref{eq:13}) by the substitution of the full gap function \(\tilde{\Delta}_a(\omega) \rightarrow \tilde{\Delta}_b(\omega)\) i.e. we replace the inner hole band gap function by the outer/larger hole band gap function, such that, we also replace all the corresponding renormalization functions i.e. \(Z_a \rightarrow Z_b\) and the related density of states.

For \(s_{++}\) symmetry, we find that the response function for the energies \(\omega < \Delta_b\) is zero over a large range and becomes non-zero only at \(\omega = 75 \,cm^{-1}\) and remains positive afterwards. This is in contrast to the behaviour of the response function given in Fig.5, for the same symmetry. Where, the function goes through the zero towards the negative peak situated near the larger gap energy i.e. \(\Delta_b\). Only for \(\mathbf{\tilde{q}}=0\), we have a response function that stays positive over the full energy range. At energies \(\omega \ge \Delta_b\), we observe that the response function peaks are shifted towards higher energies with increase in \(\mathbf{\tilde{q}}\) in both the figures. However, in Fig.7, for the \(s_{++}\) case,  there are only single positive peaks, i.e. only single mode, for all the \(\mathbf{\tilde{q}}\).
	
In the \(s_{\pm}\) case, as depicted in Fig. 7, the response function amplitude has a very large value, in fact an order of magnitude larger, than the \(s_{++}\) case in the same figure and also in comparison to the response amplitudes in Fig.5. for both the \(s_{++}\) \text{and} \(s_{\pm}\) symmetry case. 
The reason for such a behaviour is the contribution of the divergent term \(1/(\Delta_b-\omega)\) in the coherence factor \(K(\omega)\) for the \(s_{\pm}\) case, instead of a constant scalar multiple for the  \(s_{++}\) case (see Eq.(13)). 
In the region \(\omega \approx \Delta_b\), there is a large negative peak of the response function. At \(\omega > \Delta_b\) both the graphs in the upper and lower panel of Fig.7 are qualitatively similar for the increasing value of \(\mathbf{\tilde{q}}\), along with the presence of an additional mode, which is shifted towards higher values of \(\omega\), in all the cases without exception.

Although, a difference is present between both the symmetries at \(\omega \approx \Delta_b\) for this scattering; it only exists within a very narrow energy range. Hence, 
we shall confine the study to the previous case of the scattering of quasiparticles between the smaller/inner hole-like band the gap-averaged electron-like bands to study QPI. In the following, we emphasize that this robustness of the QPI response peak, with respect to various parameters, provides an ideal tool to probe the order parameter phase symmetry.
\begin{figure}[]
	\includegraphics[width=0.45\columnwidth]{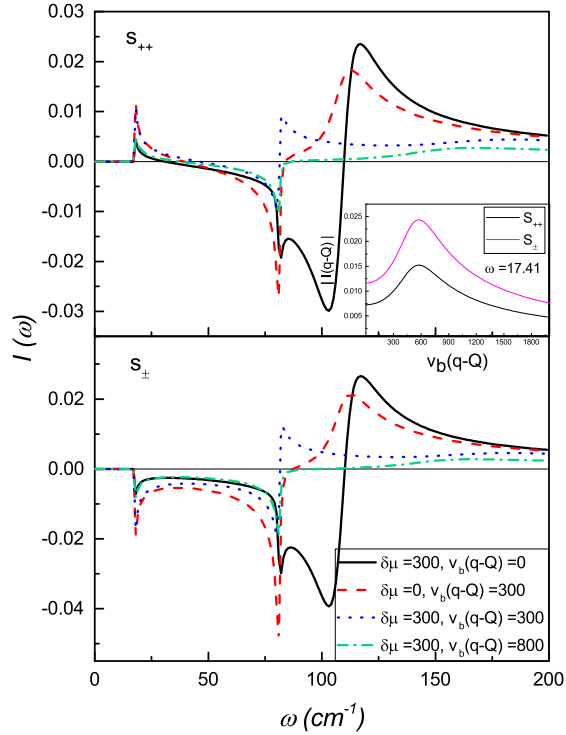}  
	\caption{The 2D plot of the QPI response function I($\protect\omega$) vs $%
		\protect\omega$, momentum $\mathbf{q}$
		and the shifted Fermi surface energy $\protect\delta\protect\mu$ for the
		strong coupling case at temperature T = 1. In the inset, the dependence of $%
		|I(\mathbf{q-Q})| $ is shown at fixed energy close to the smaller gap
		i.e. $\protect\omega \approx$ 18 $cm^{-1}$. The values of the coupling
		constants are $\protect\lambda_{aa}=0.5$, $\protect\lambda_{ab}=0.2$, $%
		\protect\lambda_{ba}=0.1$, $\protect\lambda_{bb}=3$.}
	\label{Fig.Muvfqa}
\end{figure}

In Fig. 8, the graph depicts the behaviour of the QPI response function for
very large shifted Fermi surface energy i.e. $\delta\mu$=300 and the
comparison with the case of zero $\delta\mu$ and non-zero value $\mathbf{v}%
_{b}\mathbf{\tilde{q}}$ for both the symmetry cases. The behaviour of $%
\mathbf{v}_{b}\mathbf{\tilde{q}}$ is shown by dashed curves as the momentum
vector $\mathbf{\tilde{q}}$ varies from small to large values and connects
the two order parameters on the Fermi surfaces when its of the order $(\pi)$%
. The black curve shows the behaviour of the response function for zero
momentum and large shifted Fermi surface energy. The red dashed curve for
zero $\delta\mu$ and large $\mathbf{v}_{b}\mathbf{\tilde{q}}$, shows the
difference in the two cases with a shifting of the peak that arises for $%
\omega > \Delta_b$.

For the equal values of both the parameters, the behaviour is depicted by
the blue dotted curve; where, the inverted peaks near the first and second
band gap energies are almost equal in magnitude. Finally, the green curve
shows the case for very large electron like quasiparticle momentum in
comparison to the shifted Fermi surface energy and shifted peak is shown to
be highly dispersed. The value of $\omega^*$ calculated through relation $%
\omega^* = \sqrt{ (\Delta_a^2 + \Delta_b^2 +\delta \mu^2)^2- 4\Delta_a^2
\Delta_b^2}/(2|\delta\mu|)$, for the case when $\delta\mu > \Delta_b$ is
162.03 $cm^{-1}$.

As stated previously, the most robust feature is the peak of the response
function around the first band gap energy, which does not change the sign
reversing behaviour with the change in parameters viz. $\delta\mu$, $\epsilon
$ or $\mathbf{\tilde{q}}$ in the Eq.(\ref{eq:15}). Hence, this
characteristic of the QPI response function presents itself as a very useful
feature for the probe of order parameter symmetry between the $s_{++}$ and
$s_{\pm}$ case, via the c-axis measurements from the FT-STM studies.

The inset in the upper panel of Fig. 8, depicts the strong dependence of the
magnitude of the peak on the parameter $\mathbf{\tilde{q}}$. For the perfect
nesting case, i.e. $\mathbf{q} = \mathbf{Q}$, we observe the maximum in
response function magnitude. For a fixed $\delta\mu$ and for the energy
chosen to be near $\Delta_a$, we have the experimentally tunable parameter $%
\mathbf{\tilde{q}}$ start at zero and scan over larger values. The peaks of $%
|I(\mathbf{\tilde{q}})| $ in both the symmetry cases emerge for some optimal
value of the momentum i.e. when $\mathbf{\tilde{q}}$ becomes of the order $%
\delta\mu$ (in accordance with Eq.(\ref{eq:15})). At small values of $%
\mathbf{\tilde{q}}$, this magnitude of the peaks is quite small; and hence,
to observe this experimentally, we need to find the match between the large
value of $\mathbf{\tilde{q}}$ and $\delta\mu$ to sample such behaviour
correctly.

\section{Summary \& Conclusion}

We have analysed the problem of the identification of the order parameter
symmetry for the Fe-based superconductors via the QPI measurements. For this
purpose, we have developed a theory of the quasiparticle interference in
multiband superconductors based on strong-coupling Eliashberg approach. In
particular case of a two-band system, we consider two possible pairing
symmetries the $s_{\pm}$ state, when the sign of the order parameters
changes between the hole and the electron bands and the more conventional $%
s_{++}$ state. 

The obtained results confirm the concept that the QPI is
phase-sensitive technique and may help to determine pairing symmetry in
Fe-based superconductors; and in general, could be applicable to other
multiband superconductors. We calculate energy, temperature and momentum
dependencies of the QPI response and point out qualitative differences
between the response in the $s_{\pm}$ and $s_{++}$ cases. Application of the
Eliashberg approach allows to take into account self-consistent retardation
effects due to strong coupling and to properly describe temperature
dependence of the QPI response function at various energies.
Further, we have analyzed various regimes of the Fermi surface anisotropy by
taking into account the influence of Fermi surface ellipticity.

We argued from the analysis that, in general, for $\mathbf{q} \approx \mathbf{Q}$, there are three singularities of the response function. Two of these are momentum independent (weak momentum dependence) \( \omega \approx \Delta_{a,b}(\omega)\) and the one having a strong momentum dependence. Only the momentum independent (weak momentum dependence) peak, corresponding to the lowest gap value \(\Delta_a\), may serve as a universal probe for the gap symmetry in the multiband superconductors. We emphasize that our analysis presents a convincing case in favour of the QPI measurements as a phase sensitive test of the gap symmetry for the FeBS. 
This conclusion is based on the robustness of the response function peak near the smaller gap energy and is independent of the exact nature or shape of the energy bands.

\section{Acknowledgements}

We acknowledge useful discussions with P. Hirschfeld, I.I. Mazin, D. Morr and Y.
Tanaka. This work was financially supported by the Foundation for
Fundamental Research on Matter (FOM), associated with the Netherlands
Organization for Scientific Research (NWO), by Russian Science Foundation,
Project No. 15-12-30030, and by Ministry of Education and Science of the
Russian Federation, grant 14.Y26.31.0007. DE acknowledges DFG financial support through the grant GR 3330/4-1 and financial support of VW-foundation through the grant ``\textit{Synthesis, theoretical examination and experimental investigation of emergent iron-based superconductors}".

\pagebreak

\section{Appendix}

Here, we show the 2D and 3D graphs for the response function variation
with shifted Fermi surface energy $\delta\mu$ versus the energy $\omega$
and with the electronic band ellipticity, $\epsilon$ = 0, for both the $s_{++}$
and $s_{\pm}$ cases, as discussed in the main text under sec. III(B). 

First, in Fig. 9, the trend for the response function at zero ellipticity is
presented. The response curve near the second band gap energy has a sharp
small negative peak and a broadened secondary peak as the $\delta\mu$ values
increase. The second peak shifts away from $\Delta_b$ with larger
values of shifted Fermi energy between the electron-like and hole-like
pockets and for very large $\delta\mu$ the two lower peaks become relatively
similar in strength. The positive peak around the same energy interval also
shows a shift towards $\omega > \Delta_b$ and flattens out at very
high $\delta\mu$ value. Here, again we observe that the peaks around the smaller band gap is a robust feature with respect to the variation in the parameters.

\begin{figure}[]
	\includegraphics[width=0.45\columnwidth]{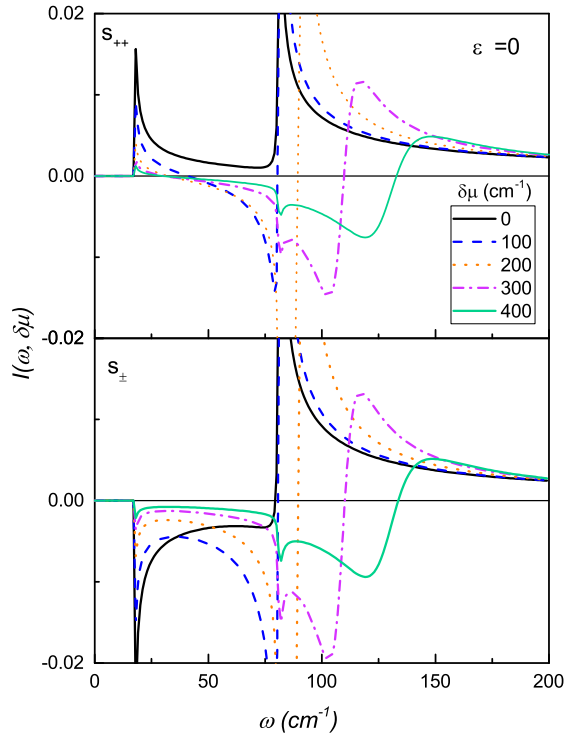}  
	\caption{The 2D plot of the QPI response function I($\protect\omega$) vs $%
		\protect\omega$ and the shifted Fermi surface energy $\protect\delta\protect%
		\mu$ for the strong coupling case at temperature T = 1 at ellipticity $%
		\protect\epsilon$ and the momentum $\tilde{q}$ =0. The values of the
		coupling constants are $\protect\lambda_{aa}=0.5$, $\protect\lambda_{ab}=0.2$%
		, $\protect\lambda_{ba}=0.1$, $\protect\lambda_{bb}=3$.}
	\label{Fig.eps0}
\end{figure}

The 3D graph in Fig. 10, shows the change in response function as we move from $\omega <
\Delta_a$ to the region $\omega > \Delta_b$. The response
function gets the inverted peak near the second band gap energy in both the
cases and there is a secondary dip that shifts towards higher energy with
increasing shifted Fermi surface energy. The shift of the second peak at $%
\omega > \Delta_b$ is observed. There is almost similar amplitude of the QPI response in both the cases with the strong coupling around the region $\omega$ = $%
\Delta_a$ for $\epsilon$ =0 case as compared to Fig. 2. For higher energies and larger
chemical potential, apart from strong peaks, we have no other distinguishing
feature for both the cases except for the QPI peak around the smaller band
gap, $\Delta_a$.

\begin{figure}
	\centering
	\subfloat[\(s_{++}\) symmetry]{{\includegraphics[width=0.45\columnwidth]{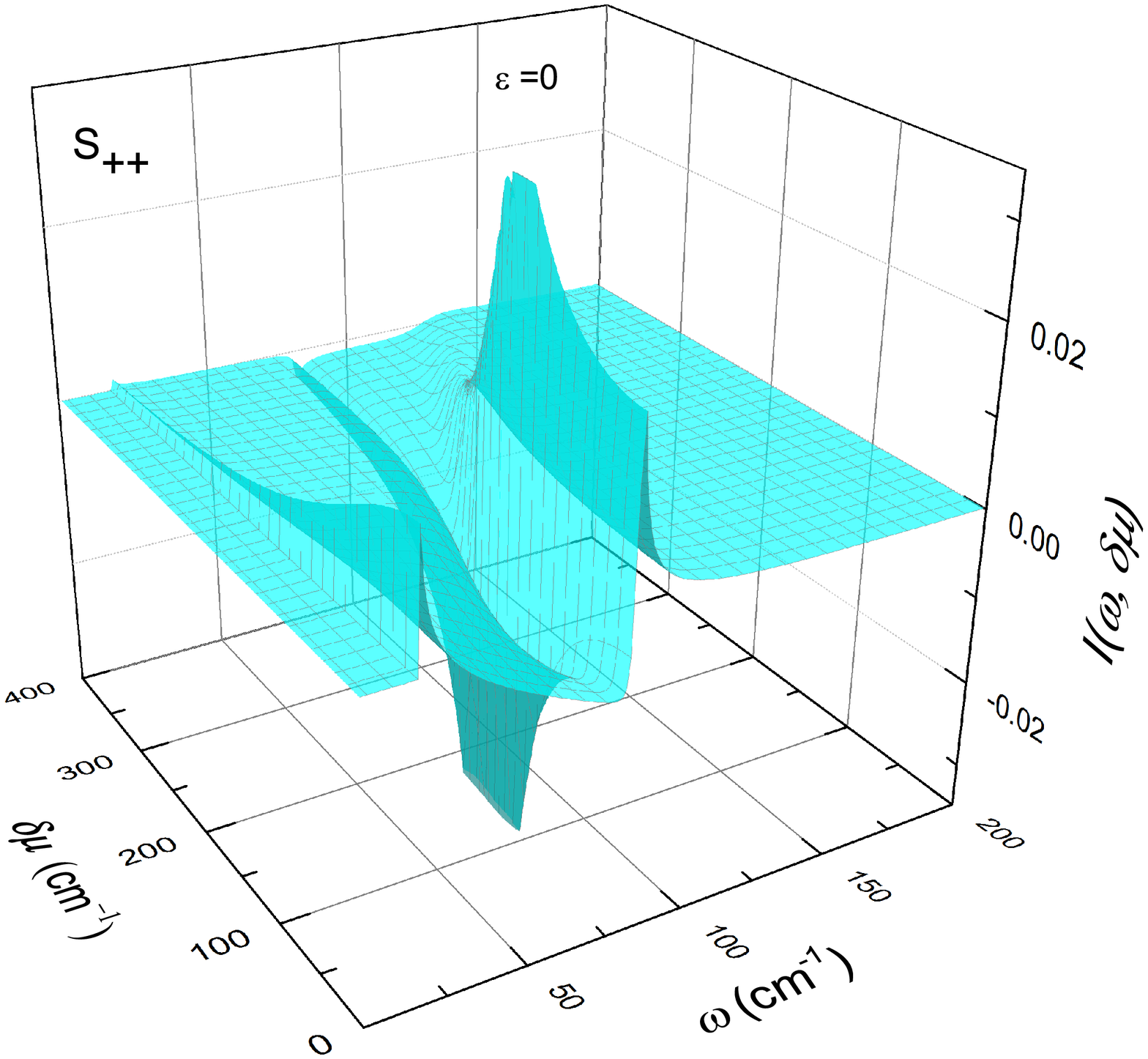} }}
	\qquad
	\subfloat[\(s_{\pm}\) symmetry]{{\includegraphics[width=0.45\columnwidth]{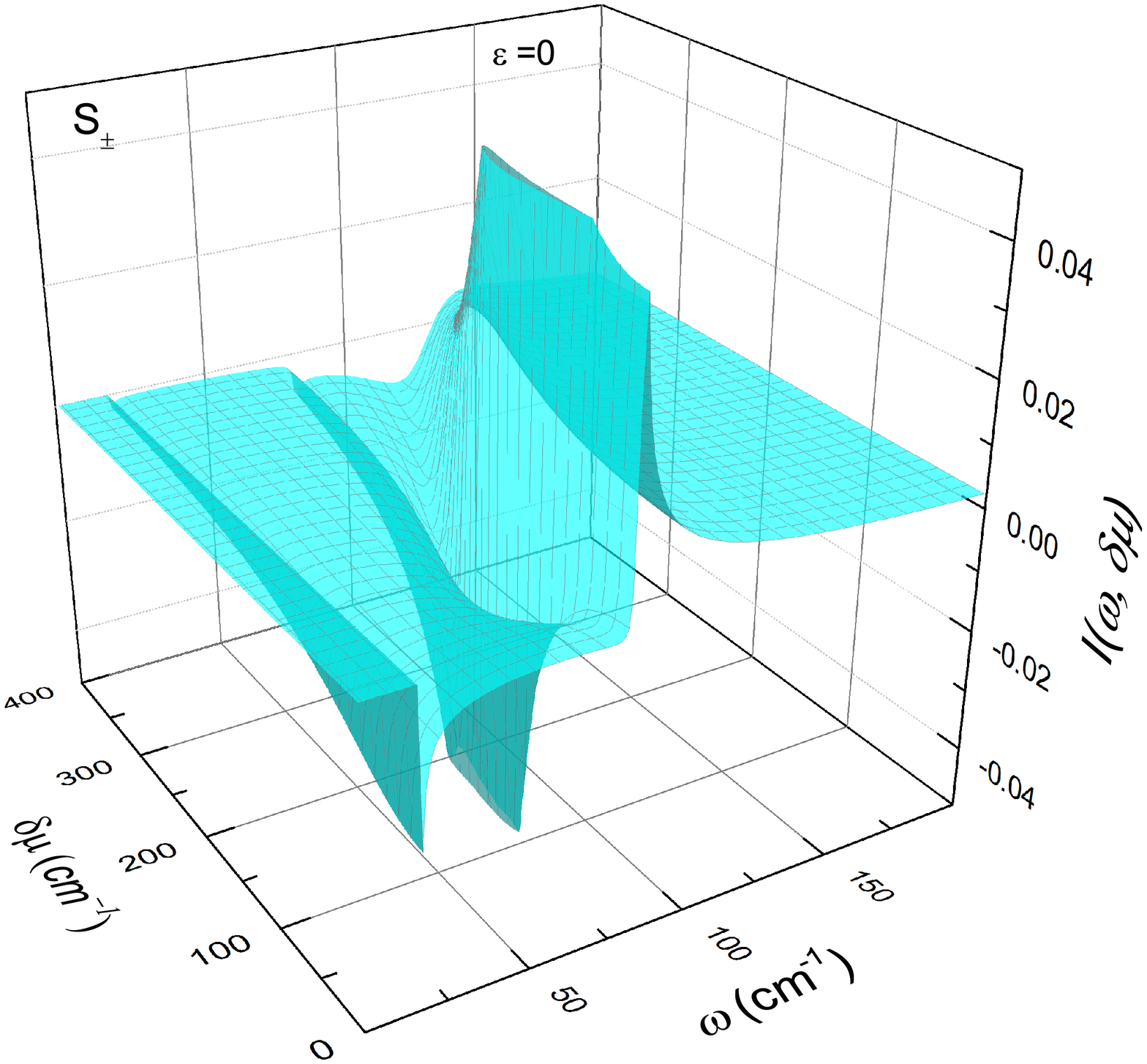} }}
	\caption{The 3D plot of the QPI response function I($\protect\omega,\protect%
		\delta\protect\mu$) vs $\protect\omega$ and the non-zero Fermi surface
		energy $\protect\delta\protect\mu$ at zero ellipticity for the strong
		coupling case at temperature T = 1 for $s_{++}$ and $s_{\pm}$. There is a
		large amplitude for the response function in the region $\protect\delta%
		\protect\mu$=[100,200] for the latter case. The values of the coupling
		constants are $\protect\lambda_{aa}=0.5$, $\protect\lambda_{ab}=0.2$, $%
		\protect\lambda_{ba}=0.1$, $\protect\lambda_{bb}=3$.}
	\label{Fig.eps1}
\end{figure}

 The effect of the relative shift of the Fermi
surface energy to a non-zero value shows that there is a rather strong
suppression of the second response peak in $s_{++}$ case as compared to the $%
s_{\pm}$ in the region $\omega\approx\Delta_b$ as compared to the
finite ellipticity case discussed below. 

\begin{figure}[]
\includegraphics[width=0.45\columnwidth]{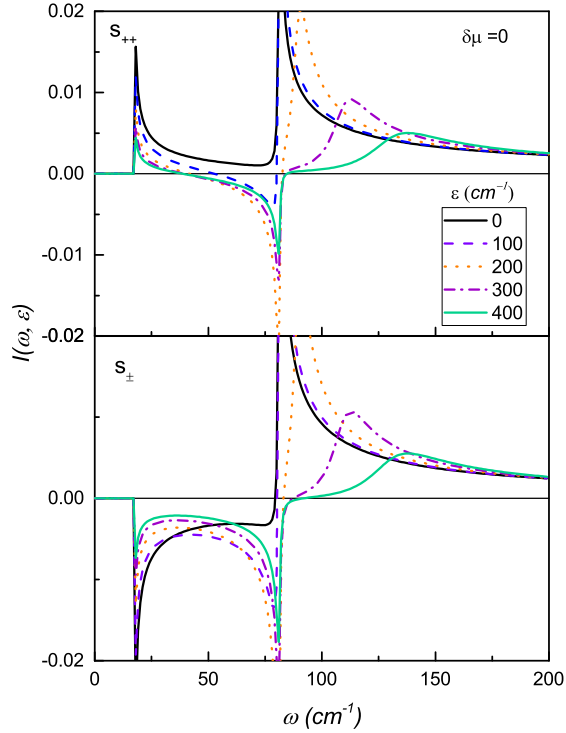}  
\caption{The 2D plot of the QPI response function I($\protect\omega$) vs $%
\protect\omega$ and the ellipticity $\protect\epsilon$ for the strong
coupling case with value of shifted Fermi surface energy $\protect\delta%
\protect\mu$ and the momentum $\tilde{q}$ =0, at temperature T = 1.  The values of the coupling constants are $\protect%
\lambda_{aa}=0.5$, $\protect\lambda_{ab}=0.2$, $\protect\lambda_{ba}=0.1$, $%
\protect\lambda_{bb}=3$.}
\label{Fig.mu0}
\end{figure}

In Figs. 11 and 12, we present the change of the response function with
variation in the band ellipticity $\epsilon$ as in Eq.(\ref{eq:14}) and setting
the shift in Fermi surface energy $\delta\mu$ = 0 with 2D and 3D graphs.  The larger
ellipticity values lead to the inversion of the peak around second band gap,
which reaches its maximum value around $\epsilon$ =200 and thereafter the
overall amplitude drops, with the positive peak dampening strongly and
shifting towards higher $\omega$ values. The peaks near the first band gap
energy are unaltered by the change of the ellipticity and hence present a
strong case for the probing of the gap symmetry based on QPI experiments. 

\begin{figure}
	\centering
	\subfloat[\(s_{++}\) symmetry]{{\includegraphics[width=0.45\columnwidth]{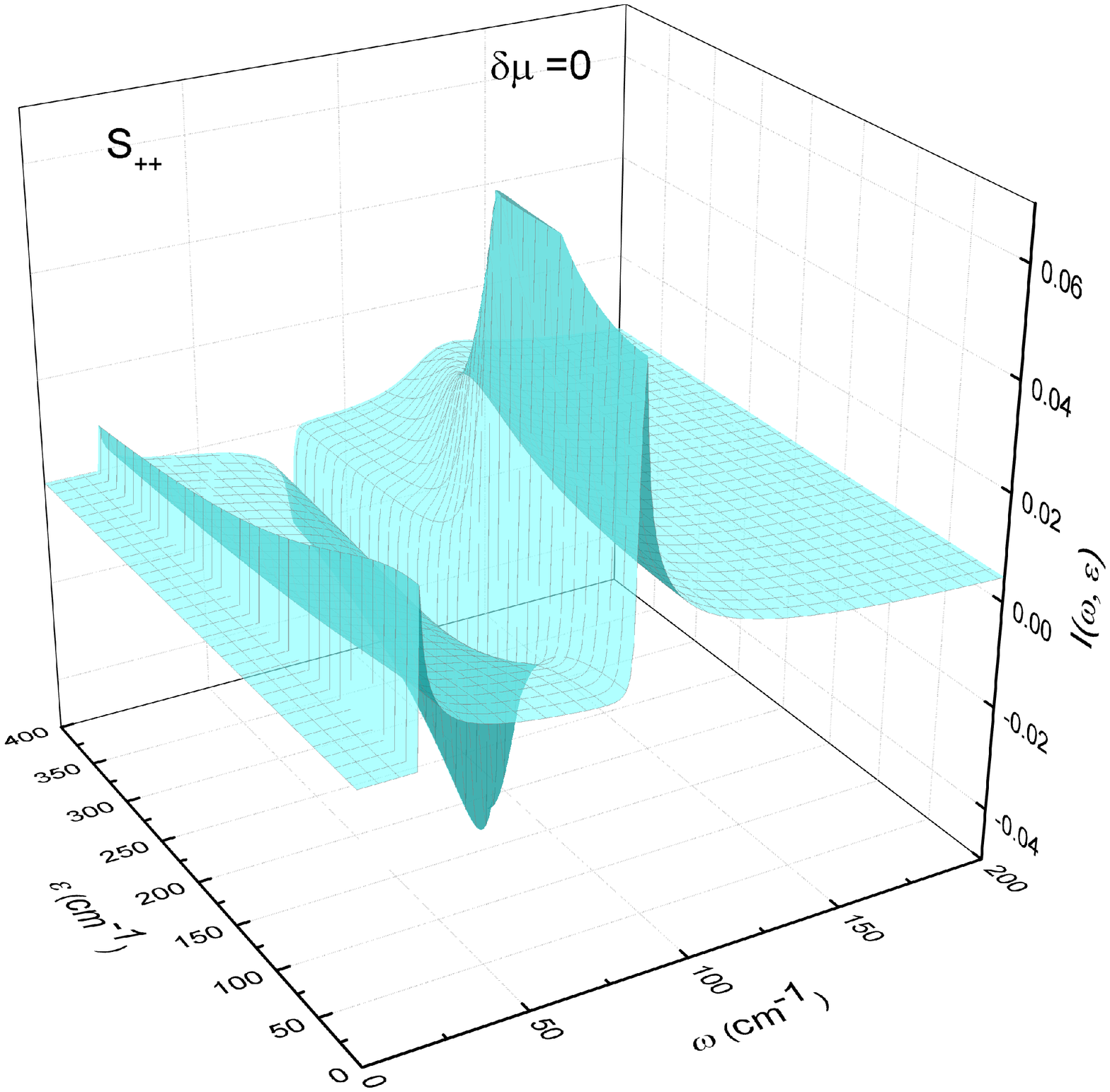} }}
	\qquad
	\subfloat[\(s_{\pm}\) symmetry]{{\includegraphics[width=0.45\columnwidth]{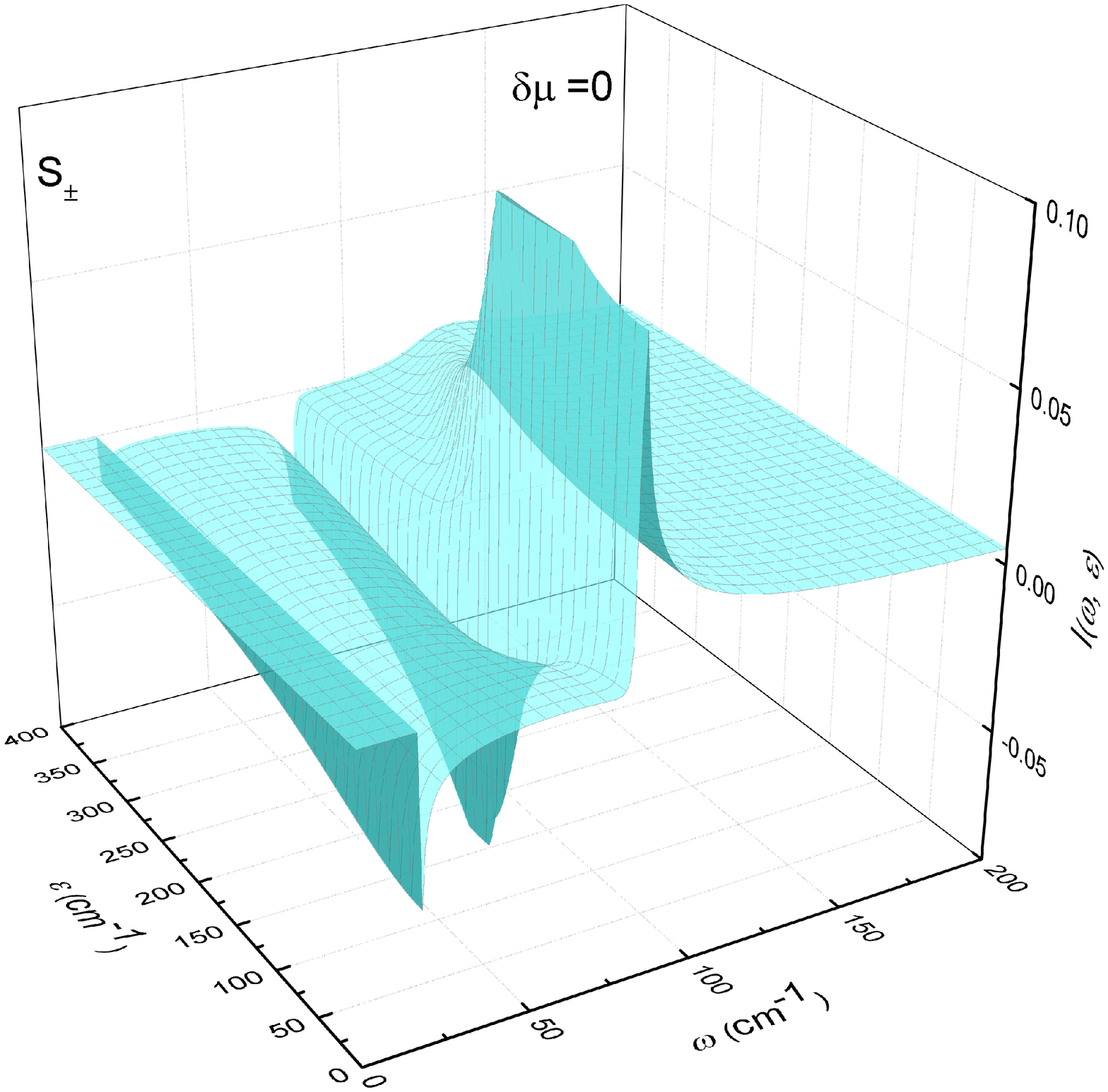} }}
	\caption{The 3D plot of the QPI response function I($\protect\omega$) vs $%
		\protect\omega$ and the ellipticity $\protect\epsilon$ for the strong
		coupling case with value of shifted Fermi surface energy $\protect\delta%
		\protect\mu$ and the momentum $\tilde{q}$ =0, at temperature T = 1. The values of the coupling constants are $\protect%
		\lambda_{aa}=0.5$, $\protect\lambda_{ab}=0.2$, $\protect\lambda_{ba}=0.1$, $%
		\protect\lambda_{bb}=3$.}
	\label{Fig.mu1}
\end{figure}

Additionally, for the energies close to the second band gap energy and with
a large $\epsilon$, the response function is negatively peaked for both the
cases and has a stronger peak around $\epsilon$ =200 with a very strongly
damping for very high ellipticity values. In both the cases, we observe the
shifting and high suppression of the positive peak towards energies $\omega
> \Delta_b$ and the negative response peak just falls off very slowly without the shift. This gains confirms our assertion that the smaller band gap peak is a promising feature that could be used as a universal tool for the pairing symmetry measurements.


\begin{thebibliography}{99}
\bibitem{Hosono2008} Y. Kamihara, T. Watanabe, M. Hirano, H. Hosono, J. Am.
Chem. Soc., \textbf{130 }, 3296, (2008).

\bibitem{Stewart11} G.R. Stewart, Rev. Mod. Phys. \textbf{83}, 1589, (2011).

\bibitem{Paglione10} J. Paglione and R.L. Greene, Nat. Phys. \textbf{6},
645, (2010).

\bibitem{Johnston10} D.C. Johnston, Adv. Phys. \textbf{59}, 803, (2010).


\bibitem{Hosono15} H. Hosono, K. Kuroki, Physica C \textbf{514} 399, (2015).

\bibitem{Mazin02} I. I. Mazin, O. K. Andersen, O. Jepsen, O.V. Dolgov, J.
Kortus, A. A. Golubov, A. B. Kuzmenko, and D. van der Marel, Phys. Rev.
Lett. \textbf{89}, 107002, (2002).

\bibitem{Cruz08} C. de la Cruz, Q. Huang, J. W. Lynn, Jiying Li, W. Ratcliff
, J. L. Zarestky, H. A. Mook, G. F. Chen, J. L. Luo, N. L. Wang \& P. Dai,
Nature, \textbf{453}, 899, (2008).

\bibitem{Hidenori09} T. Hidenori, Nature Materials \textbf{8}, 251, (2009).

\bibitem{Pratt09} D. K. Pratt, W. Tian, A. Kreyssig, J. L. Zarestky, S.
Nandi, N. Ni, S. L.  Bud'ko, P. C. Canfield, A. I. Goldman \& R. J.
McQueeney, Phys. Rev. Lett. \textbf{103}, 087001, (2009).

\bibitem{Hirsch2016} P. J. Hirschfeld, C. R. Physique \textbf{17}, 197-231,
(2016).

\bibitem{Feng2015} J-Feng Ge, Zhi-Long Liu, C. Liu, C-Lei Gao, D. Qian,
Qi-Kun Xue, Y. Liu J-Feng Jia, Nature Materials \textbf{14}, 285-289, (2015).


\bibitem{Singh2008} D. J. Singh \& M. H. Du, Phys. Rev. Lett., \textbf{100},
237003, (2008). %American Physical Society

\bibitem{Kontani2015} Y. Yamakawa \& H. Kontani, Phys. Rev. B \textbf{92},
045124, (2015).


\bibitem{Johannes2008} I. I. Mazin, D. J. Singh , M. D. Johannes \& M.H. Du
, Phys. Rev. Lett.\textbf{101}, 057003, (2008).

\bibitem{McElroy2003} McElroy et. al, Nature \textbf{422}, 592 (2003)

\bibitem{Hirschfeld2011} P.J. Hirschfeld, M.M. Korshunov, \& I.I. Mazin,
Rep. Prog. Phys. \textbf{74},124508, (2011).



\bibitem{Mazin2010} Mazin, I. I., Nature \textbf{464}, 183 (2010). 
%Superconductivity gets an iron boost.

\bibitem{Chubukov2008} A. V. Chubukov, D. V. Efremov \& I. Eremin, Phys.
Rev. B, American Physical Society, \textbf{78}, 134512 (2008).

\bibitem{Hanaguri2010} T. Hanaguri, S. Niitaka, K. Kuroki, \& H. Takagi,
Science, \textbf{328}, 474 (2010).

\bibitem{Kordyuk12} A. A. Kordyuk, Low Temperature Physics, \textbf{38},
888,(2012).


\bibitem{Kuroki8} K. Kuroki, S. Onari, R. Arita, H. Usui, Y. Tanaka, H.
Kontani \& H.Aoki, Phys. Rev. Lett. \textbf{101},087004, (2008).

\bibitem{Werner2012} P. Werner, M. Casula, T. Miyake, F. Aryasetiawan, A. J.
Millis \& S.Biermann Nature Physics \textbf{8}, 331\^{a}(2012).

\bibitem{Chen2014} X. Chen, P. Dai, D. Feng, T. Xiang \& F. C. Zhang,
National Science Review, \textbf{1}, 371 (2014).

\bibitem{Monthoux 2001} P. Monthoux \& G. G. Lonzarich, Phys. Rev. B, 
\textbf{63}, 054529 (2001).

\bibitem{Korshunov2011} P. J. Hirschfeld, M. M. Korshunov \& I. I. Mazin,
Reports on Progress in Physics, \textbf{74}, 124508 (2011).

\bibitem{Golubov2008} L. Boeri, O. V. Dolgov, and A. A. Golubov, Phys. Rev.
Lett. \textbf{101}, 026403 (2008).


\bibitem{Efremov2011} D.V. Efremov, M. M. Korshunov, O.V. Dolgov, A. A.
Golubov, \& P. J. Hirschfeld, Phys. Rev. B \textbf{84}, 180512 (2011).

\bibitem{Korshunov2014} M. M. Korshunov, D.V. Efremov, A. A. Golubov, and
O.V. Dolgov, Phys. Rev. B \textbf{90}, 134517 (2014).

\bibitem{Efremov13} D.V. Efremov, A. A.~Golubov \& O. V. Dolgov, New Journal
of Physics, \textbf{15},013002 (2013).

\bibitem{shilling2016} M. B. Schilling, A. Baumgartner, B. Gorshunov, E. S.
Zhukova, V. A. Dravin, K.V. Mitsen, D.V. Efremov, O.V. Dolgov, K. Iida, M.
Dressel, and S. Zapf, Phys. Rev. B \textbf{93}, 174515 (2016).

\bibitem{Reid2012} J. P. Reid, M. A. Tanatar, A. Juneau-Fecteau, R. T.
Gorton, S. R. de Cortret, N. Doiron-Leyraud, T. Saito, H. Fukuzawa, Y.
Kohoni, K. Kihour, C. H. Lee, A. Iyo, H. Eisaki, R. Prozorov, and L.
Taillefer, Phys. Rev. Lett. \textbf{109}, 087001 (2012).

\bibitem{Hafeiz2013} M. Abdel-Hafiez, V. Grinenko, S. Aswartham, I. Morozov,
M. Roslova, O. Vakaliuk, S. Johnston, D.V. Efremov, J. van den Brink, H.
Rosner, M. Kumar, C. Hess, S. Wurmehl, A. U. B. Wolter, B. Buechner, E. L.
Green, J.Wosnitza, P. Vogt, A. Reifenberger, C. Enss, M. Hempel, R.
Klingeler, and S. L. Drechsler, Phys. Rev. B \textbf{87} 180507 (2013).

\bibitem{Grinenko2014} V. Grinenko, D.V. Efremov, S. L. Drechsler, S.
Aswartham, D. Gruner, M. Roslova, I. Morozov, K. Nenkov, S. Wurmehl, A. U.
B. Wolter, B. Holzapfel, and B. Buechner, Phys. Rev. B \textbf{89},060504
(2014).

\bibitem{Grinenko20142} V. Grinenko, W. Schottenhamel, A. U. B. Wolter, D.V.
Efremov, S. L. Drechsler, S. Aswartham, M. Kumar, S. Wurmehl, M. Roslova,
I.V. Morozov, B. Holzapfel, B. Buechner, E. Ahrens, S. I. Troyanov, S.
Koehler, E. Gati, S. Knoener, N. H. Hoang, M. Lang, F. Ricci, and G.
Profeta, Phys. Rev. B \textbf{90}, 094511 (2014).

\bibitem{Wang2016} Q.Wang, J. T. Park, F. Y., Y. Shen, Y. Hao, Y. Pan, J.
Lynn, A. Ivanov, S. Chi, M. Matsuda, H. Cao, R. J. Birgenau, D.V. Efremov,
and J. Zhao, Phys. Rev. Lett. \textbf{116}, 197004 (2016).

\bibitem{Golubov2013} A. A. Golubov \& I. I. Mazin, Applied Physics Letters, 
\textbf{102}, (2013).

\bibitem{Mazin2009} D. Parker \& I. I. Mazin, Phys. Rev. Lett. \textbf{102},
227007, (2009).

\bibitem{Burmistrova2015} A. V. Burmistrova, I. A. Devyatov, A. A. Golubov,
K. Yada, Y. Tanaka, M. Tortello, R. S. Gonnelli, V. A. Stepanov, X. Ding, H.
H. Wen, \& L. H. Greene, Phys. Rev. B \textbf{91}, 214501, (2015).

\bibitem{Golubov2009} A. A. Golubov, A. Brinkman, Yukio Tanaka, I. I. Mazin,
\& O. V. Dolgov, Phys. Rev. Lett. \textbf{103}, 077003, (2009).



\bibitem{Crommie1993} M.F. Crommie et. al, Nature, \textbf{363}, 524- 527
(1993).

\bibitem{Kanisawa2001} K. Kanisawa, M. J. Butcher, H. Yamaguchi \& Y.
Hirayama, Phys. Rev. Lett., \textbf{86}, 3384 (2001).

\bibitem{Hoffman2002} J. E. Hoffman, K. McElroy, D.H. Lee, K. M. Lang, H.
Eisaki, S. Uchida, J. C. Davis, Science, \textbf{297}, 1148 ( 2002).

\bibitem{Howald2003} C. Howald, H. Eisaki, N. Kaneko, M. Greven, and A.
Kapitulnik, Phys. Rev. B, \textbf{67}, 014533 (2003).

\bibitem{Hanaguri2007} T. Hanaguri, Y. Kohsaka, J. C. Davis, C. Lupien, I.
Yamada, M. Azuma, M.  Takano, K. Ohishi, M. Ono \& H. Takagi, Nature Physics
, \textbf{3}, 865 (2007).

\bibitem{Scalapino2003} L. Capriotti, D. J. Scalapino \& R. D. Sedgewick,
Phys. Rev. B, \textbf{68},014508 (2003).

\bibitem{Coleman09} M. Maltseva and P. Coleman, Phys. Rev. B \textbf{80},
144514 (2009).

\bibitem{Skyora2011}S. Sykora and P. Coleman, Phys. Rev. B \textbf{84}, 054501 (2011).

\bibitem{Hirschfeld15} P.J. Hirschfeld, D. Altenfeld, I. Eremin, I.I. Mazin,
Phys. Rev. B \textbf{92}, 184513 (2015).

\bibitem{Scalapino2012} D. J. Scalapino, Rev. Mod. Phys., \textbf{84}, 1383
(2012).

\bibitem{Scalapino66} D. J. Scalapino, J. R. Schrieffer \& J. W. Wilkins,
American Physical Society, \textbf{148}, 263 (1966).

\bibitem{Allen82} P. B. Allen and B. Mitrovic, 'Theory of Superconducting
Tc. : '$\mathit{Solid \, State \, Physics}$', F. Seitz, D. Turnbull, and H.
Ehrenreich, eds. (Academic, New York, 1982) pp.1-92

\bibitem{Carbotte90} J. P. Carbotte, Rev. Mod. Phys., \textbf{62}, 1027 (
1990).

\bibitem{Marsiglio2008} F. Marsiglio \& J. P. Carbotte in J. Bennemann, K.
\& Ketterson, J. (Eds.): Electron-Phonon Superconductivity, $\mathit{\
Superconductivity}$, Springer Berlin Heidelberg, 73 (2008).

\bibitem{Scalapino69} D.J. Scalapino 1969 in $\mathit{\ Superconductivity \,}
$ ed. by Parks R D (New York: Marcel Dekker) Chap {0} 449

\bibitem{Parker08} D. Parker, O.V. Dolgov, M.M. Korshunov, A.A Golubov, and
I.I. Mazin, Phys. Rev. B \textbf{78} 134524 (2008).

\bibitem{Maksimov82} E.G. Maksimov and D.I. Khomskii 1982 \textit{High
Temperature Superconductivity } ed. Ginzburg V L and Kirzhnitz D (New York:
Consultant Bureau).

\bibitem{Vonsovskii82} S.V. Vonsovskii, Yu A Izjumov, E.Z. Kurmaev 1982 
\textit{Superconductivity of transition metals: their alloys and compounds}
(Berlin: Springer 1982).

\bibitem{Inosov09} D.S. Inosov, J.T. Park, P. Bourges, D.L. Sun, Y. Sidis,
A. Schneidewind, K. Hradi, D. Haug, C.T. Lin, B. Keimer and V. Hinkov 
\textit{et al.} Nature Physics, \textbf{6} 178 (2009).

\bibitem{Popo2010} P. Popovich, A. V. Boris, O. V. Dolgov, A. A. Golubov, D.
L. Sun, C. T. Lin, R. K. Kremer, and B. Keimer Phys. Rev. Lett. \textbf{105}%
, 027003 (2010).

\bibitem{Charnukha2010} A. Charnukha, O. V. Dolgov, A. A. Golubov, Y.
Matiks, D. L. Sun, C. T. Lin, B. Keimer, and A. V. Boris Phys. Rev. B 
\textbf{84}, 174511 (2011).

\bibitem{Charnukha2014} A. Charnukha, J. Phys. Condens. Matter \textbf{26},
253203 (2014).

\bibitem{Dolgov2011} A.A. Golubov, O.V. Dolgov, A.V. Boris, et al. JETP
Lett. \textbf{94} 333 (2011).

\bibitem{Dolgov2009} O. V.Dolgov, I. I.Mazin, D. Parker \& A. A. Golubov,
Phys. Rev. B, \textbf{79},060502 (2009).
\end{thebibliography}
\end{document}